\documentclass[10pt,a4paper,english,oneside]{article}
\usepackage[english]{babel}
\usepackage[utf8]{inputenc}
\usepackage[margin=1.5in]{geometry}

\usepackage{enumerate}
\usepackage{graphicx}
\usepackage{xcolor}
\usepackage{amsmath,amsfonts}
\usepackage{comment}
\usepackage{algorithm}
\usepackage{algpseudocode}

\usepackage[noblocks]{authblk}

\usepackage[breaklinks]{hyperref}
\hypersetup{
    colorlinks,
    linkcolor={red!80!black},
    citecolor={green!70!black},
    urlcolor={blue!80!black}
}

\usepackage[babel]{csquotes}
\usepackage[style=numeric-comp,
            backend=biber,
            mcite=true,
            doi=false,
            url=false,
            isbn=false,
            eprint=false,
            giveninits=true,
            sorting=none]{biblatex}
\bibliography{refs.bib}

\newbibmacro{string+doiurlisbn}[1]{%
  \iffieldundef{doi}{%
    \iffieldundef{url}{%
      \iffieldundef{isbn}{%
        \iffieldundef{issn}{%
          #1%
        }{%
          \href{http://books.google.com/books?vid=ISSN\thefield{issn}}{#1}%
        }%
      }{%
        \href{http://books.google.com/books?vid=ISBN\thefield{isbn}}{#1}%
      }%
    }{%
      \href{\thefield{url}}{#1}%
    }%
  }{%
    \href{http://dx.doi.org/\thefield{doi}}{#1}%
  }%
}
\DeclareFieldFormat{title}{\usebibmacro{string+doiurlisbn}{\mkbibemph{#1}}}
\DeclareFieldFormat[article,incollection,inproceedings,inbook]{title}%
    {\usebibmacro{string+doiurlisbn}{\mkbibquote{#1}}}

\newcommand{\Order}[1]{\mathcal{O}\hspace{-.5mm}\left( #1 \right)}
\newcommand{\bJ}{\mathbf{J}}
\newcommand{\calN}{\mathcal{N}}
\newcommand{\expected}[1]{\left\langle #1 \right\rangle}

\newcommand{\rmd}{\mathrm{d}}
\newcommand{\diff}{\mathrm{d}}
\def\mean#1{\mathinner{\overline{#1}}}

\DeclareMathOperator{\Var}{Var}

\DeclareMathOperator{\sgn}{sgn}

\DeclareMathOperator{\erfc}{erfc}

\newcommand{\bSigma}{\boldsymbol{\Sigma}}
\newcommand{\p}[1]{\left( #1 \right)}

\newcommand{\br}{\mathbf{r}}
\newcommand{\rme}{\mathrm{e}}
\newcommand{\calZ}{\mathcal{Z}}

\newcommand{\calC}{\mathcal{C}}

\newcommand{\rmi}{\mathrm{i}}

\newcommand{\dblambda}{\underline{\boldsymbol{\lambda}}}
\newcommand{\dbx}{\underline{\boldsymbol{x}}}
\newcommand{\bA}{\mathbf{A}}

\newcommand{\bx}{\boldsymbol{x}}
\newcommand{\hP}{\hat{P}}
\newcommand{\calD}{\mathcal{D}}
\newcommand{\blambda}{\boldsymbol{\lambda}}

\newcommand{\calH}{\mathcal{H}}

\begin{document}

\title{Satisfiability transition in asymmetric neural networks}
\author[1]{Fabián Aguirre-López}
\author[1]{Mauro Pastore}
\author[1]{Silvio Franz}
\affil{\small \textit{Université Paris-Saclay, CNRS, LPTMS, 91405 Orsay, France}}

\date{\vspace{-5ex}}

\maketitle
\begin{abstract}
    Asymmetry in the synaptic interactions between neurons plays a crucial role in determining the memory storage and retrieval properties of recurrent neural networks. In this work, we analyze the problem of storing random memories in a network of neurons connected by a synaptic matrix with a definite degree of asymmetry. We study the corresponding satisfiability and clustering transitions in the space of solutions of the constraint satisfaction problem associated with finding synaptic matrices given the memories. We find, besides the usual SAT/UNSAT transition at a critical number of memories to store in the network, an additional transition for very asymmetric matrices, where the competing constraints (definite asymmetry vs. memories storage) induce enough frustration in the problem to make it impossible to solve. This finding is particularly striking in the case of a single memory to store, where no quenched disorder is present in the system.
\end{abstract}

\section{Introduction}

Models of recurrent neural networks where neurons are coupled by non-symmetric interactions were introduced in the past to overcome limitations of classical models of associative memory, as the renowned Hopfield network~\cite{hopfield1982}. In this setting, a large number of binary units representing two-states (on/off, $\pm 1$) neurons are nodes of a weighted directed graph; the edges of the graph stand for excitatory/inhibitory synapses, depending on their sign. At a certain time step, each neuron is activated (deactivated) if the weighted contribution from all its neighbours at the previous time is positive (negative), with a simultaneous update. At variance with the simpler scenario of feed-forward neural networks, the graph contains feedback loops that cannot be unrolled on a tree-like structure. As such, the network exhibits a non-trivial dynamics, whose attractors (fixed points and limit cycles) in the configuration space of the neurons can be interpreted as memories: whenever the system is sufficiently close to one of them (\textit{i.e.}, in its basin of attraction), it can recollect the corresponding configuration starting from incomplete information, following the temporal evolution~\cite{coolen2005theory}.

The need to consider generic asymmetric graphs of synapses was already suggested in the seminal paper by Hopfield~\cite{hopfield1982}. As noted in some of the early physical literature on the subject~\cite{feigelman1986,hertz1986,parisi1986asym,derrida1987,kree1991}, the motivation to do so is twofold. First of all, the symmetric interaction between neurons is not a realistic hypothesis to model biological neural networks~\cite{li1999,ko2011,esposito2014,leonetti2020}: in the brain, neurons are connected through both unidirectional and bidirectional synapses, which can be excitatory or inhibitory. Moreover, beside the interest strictly from the field of neuroscience, the study of dynamical systems with non-reciprocal couplings is considered crucial in the much broader field of out-of-equilibrium statistical mechanics and its multidisciplinary applications, as pointed out in the context of non-equilibrium critical phenomena~\cite{vitelli2021}, dynamical systems~\cite{lacroix_a_chez_toine2022}, replicators models~\cite{opper1992,galla2006}, multi-agents economy~\cite{bouchaud2013,moran2019}, ecology~\cite{bascompte2006,allesina2012,galla2018,roy2019,altieri2021}, reporting only a few examples. In these cases, the steady states of the system in the long time limit do not correspond in general to minima of an equilibrium Hamiltonian and the dynamics is irreversible.

In this paper, we will carry on a study about neural networks with definite asymmetry, as originally proposed in~\cite{gardner1989phase} by Gardner, Gutfreund and Yekutieli (GGY in the following). For a recent historical account on this work see \cite{gutfreund2021}.
In their approach, the problem of storing memories in an asymmetric network is formulated as a constraint satisfaction problem (CSP): an instance of the problem is given by a set of random memory patterns extracted from a certain probability distribution, and the question is to find how many synaptic matrices \emph{with given asymmetry} can define networks with those memories as fixed points. Increasing the number of memories to store, the number of matrices satisfying the constraints (the synaptic volume of solutions) is expected to shrink up to a certain critical value, usually called \emph{storage capacity} of the CSP, where no solution can be found. At this point, a SAT/UNSAT transition occurs.

This point of view on the problem of associative memory is interesting because it is substantially alternative and complementary to most of the works in the field from the recent past. Indeed, following Ref.~\cite{folli2018,hwang2019} and borrowing the glossary from neuroscience, we can say that the physical literature on the dynamics of the Hopfield model (and its generalization to generic asymmetry) is mainly divided between {a ``connectionist'' approach~\cite{sompolinsky1985,parisi1986asym,feigelman1986,hertz1986,hatchett2004,brunel2016,leuzzi2022} and an ``innate'' approach~\cite{crisanti1987,gutfreund1988asym,krauth1988,crisanti1993}: in the first case, the memories are stored in the network updating the synaptic weights via learning rules, such as the Hebb's rule~\cite{hebb1949} (or its slight modification to introduce asymmetry), the BCM's rule~\cite{BCM1982}, the Storkey's rule~\cite{storkey1997} (among the so-called biologically-plausible rules), or methods based on gradient descent~\cite{tolmachev2020}; in the second case, weights extracted from a certain distribution represent a brain born with a random connectome.} In both cases, progress has been achieved very recently, arousing a renewed interest in the field. 

From the ``innate'' side, for example, results on the number of limit cycles in the neurons dynamics in random asymmetric fully-connected~\cite{hwang2019} and diluted~\cite{hwang2020} networks have been obtained, via the evaluation of their complexity for large number of neurons. Even more recently~\cite{ventura2021}, in the ``connectionist'' paradigm two different dynamics of learning have been compared, unveiling a link between the symmetry of the network and the ``unlearning'' process~\cite{hopfield1983unlearning,fachechi2018dreaming} that has been proposed to improve the retrieval performance of recurrent neural networks, in analogy with the role that dream sleep~\cite{crick1983} is supposed to have in biological brains.

To clarify the difference in points of view between the GGY approach and the great majority of the literature on Hopfield networks, and to clear the field from possible misunderstandings, let us state precisely what we mean with ``storage capacity'' in the context of this paper: it is the maximum number of random uncorrelated memories for which the CSP formulated above still admits solutions. This meaning, borrowed from~\cite{gardner1987perceptron}, is different with respect to the one of~\cite{hopfield1982}, where the term is used to define the critical number of memories between a retrieval phase in the dynamics of neurons, where the planted fixed points have a finite basin of attraction, and the ``blackout catastrophe'' where the dynamics does not converge to them. However, the CSP storage capacity is an obvious upper bound of the critical point where the blackout catastrophe occurs.

We can say that, despite the recent progress in the understanding of the dynamical features of these models, the problem of assessing their storage capacity in the CSP sense remains quite open and rather obscure. In~\cite{gardner1989phase} indeed, where this CSP is introduced, at least two unsettling facts are pointed out: 
    \begin{enumerate}[(i)]
    \item The usual mean field approach seems to work only in a strongly diluted model where each of the $N$ neurons is at most connected to $C\sim O(\log N)$ neighbours. In the fully connected case $C\sim O(N)$, the series of cumulants of the synaptic volume with respect to the probability distribution of random memories does not converge due to loops contributing at any order in $1/N$ (this view is disputed in~\cite{theumann1996}).

    \item For high degree of asymmetry, the CSP shows a SAT/UNSAT transition with unclear clustering properties: the volume of solutions shrinks to 0 as the number of random memories increases, but at the transition different solutions with non-trivial overlaps seem to disappear discontinuously.
    \end{enumerate}

The main contribution of the present paper is to clarify the nature of the satisfiability transition mentioned in point (ii) above. Through a cavity analysis on a minimal model with a single memory pattern to store and a soft constraint on the asymmetry, and reconsidering the replica analysis of the original GGY problem, we will be able to conclude that the unusual behaviour of the system at the transition is not related to the formation of clusters in the space of solutions, but it is due to conflicting constraints (definite asymmetry vs. memory patterns) that, at the critical curve, induce perfect anti-correlation on the solutions of the problem.

The paper is organized as follows: in Sec.~\ref{sec:model}, we introduce the CSP studied by GGY. In Sec.~\ref{sec:cavity} we report a full analysis using the cavity method on a modified model, where only one memory pattern has to be stored in a network with finite connectivity and soft constraints on the asymmetry of the graph; no quenched disorder is present in this problem (except for the inessential fact that the model is formulated on an instance of a random regular graph, see Appendix~\ref{app:opn-replica}), but still we can identify an asymmetry-driven SAT/UNSAT transition with features similar to the one found by GGY. In Sec.~\ref{sec:replica} we go back to the many-patterns and hard-constraints case; a stability analysis in the replica framework (see also Appendix~\ref{app:replica}) makes us rule out the possibility of the occurrence of a clustering transition in the space of solutions, associated to replica symmetry breaking. In Sec.~\ref{sec:discussion} we draw our conclusions, via a comparison between the two models analyzed.

\section{CSP for recurrent neural networks with definite asymmetry}
\label{sec:model}

The dynamics of a recurrent neural network is implemented by the rule
\begin{equation}
    S_i(t+1) = \sgn\left[\sum_{j\in \partial i} J_{ij} S_j (t) \right]\,,
    \label{eq:dynamics}
\end{equation}
where the variable $S_i(t) = \pm 1$ indicates the state (on/off) of the $i$th neuron at time $t$. The matrix $\bJ$ models the network of neuronal interactions: the entry $J_{ij}$ is the synaptic weight from neuron $i$ to neuron $j$, whose modulus and sign represent, respectively, the strength and the excitatory/inhibitory nature of the interaction.

The neural network model is set on a simple $C$-regular graph, $G= (V,E)$, with node set $V = \{1,\dots,N\}$ {(where $N$ is the number of neurons)} and edge set $E$: each neuron $i{\in V}$ lives on a node of the graph and is connected to $C$ other neurons; {with $\partial i$ we denote the set of neurons $j$ connected to neuron $i$ (that is, $(i,j)\in E$ if $j\in\partial i$) and} $|\partial i| = C$; the fully-connected case is recovered for $C\to N-1$. In the following we will consider only symmetrical diluted models, for which the adjacency matrix of the graph is symmetric {(that is, if $(i,j)\in E$ then $(j,i)\in E$), and no self-connections (Hopfield network)}.
Each row of the matrix $\bJ$ is normalized on the sphere of radius $\sqrt{C}$. Moreover, the synaptic matrix is required to have a \emph{structure}, that is a certain degree of correlation between entries. Following~\cite{gardner1989phase}, we will consider matrices with a level of asymmetry fixed as a function of a parameter $\eta$, interpolating between perfectly symmetric and perfectly anti-symmetric matrices. The normalization and the definite-asymmetry conditions on the synaptic matrix $J_{ij}$ are implemented by the constraints
\begin{equation}
    \sum_{j \in \partial i} J_{ij}^2 = C\,, \qquad \sum_{j\in \partial i} J_{ij} J_{ji} = \eta C\,, \qquad \forall\,i = 1,\cdots,N\,,
    \label{eq:symm_constr}
\end{equation}
where $-1\le \eta \le 1$ (for $\eta = \pm 1$ the matrix is, respectively, symmetric/anti-symmetric). The spectral properties of this random matrix ensemble, in the fully-connected thermodynamic limit $C \to N - 1 \to \infty$, have been investigated in~\cite{sommers1988spectral}. 

An \emph{associative memory} of the network is a configuration $\xi \in \{\pm 1\}^N$ of the neurons which is a fixed point of the dynamical rule~\eqref{eq:dynamics}. Indeed, when the network is in a configuration $S(t) = \xi + \epsilon(t)$ close enough to the fixed point, \emph{i.e.} in its basin of attraction, the temporal evolution according to the dynamical rule~\eqref{eq:dynamics} converges to $\xi$ by definition, allowing the reconstruction of the full information stored in the fixed point starting from its partial knowledge.

The condition for $P = \alpha C$ configurations $\xi^\mu$ to be fixed points of this dynamics is given by
\begin{equation}
    \xi^\mu_i \sum_{j \in\partial i} J_{ij} \xi^\mu_j > \sqrt{C}\kappa\,, \qquad \forall\, \mu = 1,\cdots , P = \alpha C\,;\quad \forall\,i = 1,\cdots,N\,,
    \label{eq:pattern_constr}
\end{equation}
where $\kappa$ is a positive \emph{margin}. The fractional synaptic volume, counting how many matrices from the ensemble~\eqref{eq:symm_constr} satisfy the constraints~\eqref{eq:pattern_constr}, is defined as
\begin{multline}
    V_P =  V_0^{-1} \int \Biggl[\prod_{(i,j)\in E} \diff J_{ij}\diff J_{ji}\Biggr] \prod_{i\in V} \delta\Biggl(\sum_{j\in  \partial i} J_{ij}^2 - C\Biggr)\,\delta\Biggl(\sum_{j\in \partial i} J_{ij}J_{ji} - \eta C\Biggr)\\
    \times \prod_{\mu=1}^P \theta\Biggl(\xi^\mu_i \sum_{j\in \partial i} \frac{J_{ij} \xi^\mu_j}{\sqrt{C}} - \kappa \Biggr)\,,
\label{eq:volume}
\end{multline}
where the normalization constant is given by
\begin{equation}
\begin{aligned}
    V_0 &= \int \Biggl[\prod_{(i,j)\in E} \diff J_{ij}\diff J_{ji}\Biggr] \prod_{i\in V} \delta\Biggl(\sum_{j\in \partial i} J_{ij}^2 - C\Biggr)\,\delta\Biggl(\sum_{j\in \partial i} J_{ij}J_{ji} - \eta C\Biggr)\\
    & \sim \exp\left[{\frac{ CN}{2} \left( \log 2\pi+1\right) + \frac{CN}{4} \log(1-\eta^2) }\right]\,.
    \end{aligned}
\end{equation}
The patterns $\xi^\mu$ are quenched disordered variables taken independently and uniformly on the vertices of the hypercube $\{\pm 1\}^N$: their physical interpretation is that of random independent memories stored in the network, in the sense explained above. For $P=1$, that is a single pattern to store, we will call the model \emph{one pattern network} (OPN), following~\cite{krauth1988} (see Sec.~\ref{sec:cavity} for more details); we will call the generic case $P>1$ \emph{many patterns network} (MPN), and go back to it in Sec.~\ref{sec:replica}.

{
Note that biological neural networks are known to respect with a good approximation Dale's principle~\cite{eccles1976}, according to which all the outgoing synaptic weights from a certain neuron are of the same nature (all positive or all negative: in the first case the neuron is called excitatory, in the second case inhibitory); including these constraints on the elements of the matrix $\bJ$ is outside of the scope of the present paper, but could represent an interesting follow-up motivated by neuroscience and artificial neural networks applications (see, for example,~\cite{marti2008,esposito2014,brunel2016,cornford2021}).
}

\section{One pattern network\label{sec:cavity}}

The replica calculation performed in \cite{gardner1989phase} revealed a special SAT/UNSAT transition at a value of the asymmetry $\eta_{c}(\alpha, \kappa)$ where different solutions of the problem with non trivial overlaps seem to disappear discontinuously. In the Sec.~\ref{sec:replica} we will review these results and we will show that this phenomenon is not related to a preceding clustering transition in the space of solutions, at least under the original assumptions used by GGY. In this section instead, we will introduce a minimal model that also shows a satisfiability transition at a critical value of $\eta$. Solving it through the cavity method will allow us to understand the physical nature of this transition.

A very important assumptions behind the GGY calculation is that the matrices $J_{ij}$ were taken defining diluted networks of connectivity $C \sim \Order{\log N}$. The technical reason for this assumption was the need to justify a truncation in the expansion in cumulants of the synaptic volume~\eqref{eq:volume} (see~\cite{gardner1989phase,theumann1996} for more details). This suggests studying the finite connectivity case $C \sim \Order{1}$ in itself and then taking the limit $C\to\infty$ after the limit $N\to\infty$. In order to make some progress analytically, we will focus in this section on a model with a soft constraint on the asymmetry. While we expect this model to be exactly equivalent to the one presented in Sec.~\ref{sec:model} only at  $C\to\infty$, we will show that its phenomenology at finite $C$ is still reminiscent of what is observed in the original model. This will provide a physical interpretation on the peculiar nature of the satisfiability transition in these CSPs with asymmetry, as we will discuss in Sec.~\ref{sec:discussion}. 

We will focus our attention on the one pattern network model (soft OPN) with finite C, obtained by setting $P=1$, as in~\cite{krauth1988}. After a gauge transformation, the OPN can always be reduced to the $\xi_i = 1$ case: it is a model where no disorder is present. Nevertheless, we will still find a critical value of asymmetry, $\eta_c$ \eqref{eq:asymptotic-moments}, at which the volume of solutions vanishes. In this model it is clear that the mechanism behind the SAT/UNSAT transition is the complete anti-correlation between $J_{ij}$ and $J_{ji}$, \eqref{eq:asymptotic-moments}. We can also show that the volume of solutions is shrinking not to a point, but to a manifold defined by a specific set of equations, \eqref{eq:marginal-equations}. Notice that finite $C$ implies less degrees of freedom, making it a harder CSP than $C = \Order{\log N}$. For this reason, one should not consider the OPN with soft asymmetry an overly-simplified version of the MPN described by GGY, but rather the minimal model exhibiting the same phenomenology.

\subsection{Cavity approach for the soft OPN}

In order to study the Gardner volume for the one pattern network with a cavity approach, we need to write our problem in terms of a graphical model. We construct it in the following way. For a fixed instance of a random regular graph, $G = (V,E)$, of connectivity $C$,  we define a pair of variables on each edge. We assume $|V| = N$, and $|E| = NC/2$, so that we have a total of $NC$ degrees of freedom. We denote the complete set of variables by $\bJ$ and for the variables at each edge we introduce the following notation:
\begin{align}
\label{eq:soft-ensemble}
    \bJ_{ij} ={} &  (J_{ij},J_{ji}) \;\; \textrm{ for } i<j \textrm{ such that }(i,j) \in E
\end{align}

We define the probability measure $\rho(\bJ)$ over all the variables, consisting of a product of zero-mean Gaussian distributions, $\calN_\lambda$, with correlation $\lambda$ between edge variables, truncated by a product of the pattern constraints on each row. In this way, we can use the parameter $\lambda$ to control the correlation between the $J_{ij}$ and $J_{ji}$ while simultaneously satisfying all the constraints. In formula,
\begin{equation}
\label{eq:opn-model}
    \rho(\bJ) = \frac{1}{Z} \cdot \prod_{(i,j)\in E} \calN_{\lambda}(\bJ_{ij})\cdot \prod_{i \in V} \theta(\sum_{j\in\partial i}J_{ij} - \kappa \sqrt{C})\,,
\end{equation}
where
\begin{equation}
    \calN_{\lambda}(\bJ_{ij}) = \calN(J_{ij},J_{ji}| \boldsymbol{0}, \boldsymbol{\Sigma})\,, \qquad
    \boldsymbol{\Sigma} = \begin{pmatrix}
    1     & \lambda \\
    \lambda     & 1 
    \end{pmatrix}.
\end{equation}

\begin{figure}
	\centering
	\includegraphics[width = 0.98\textwidth]{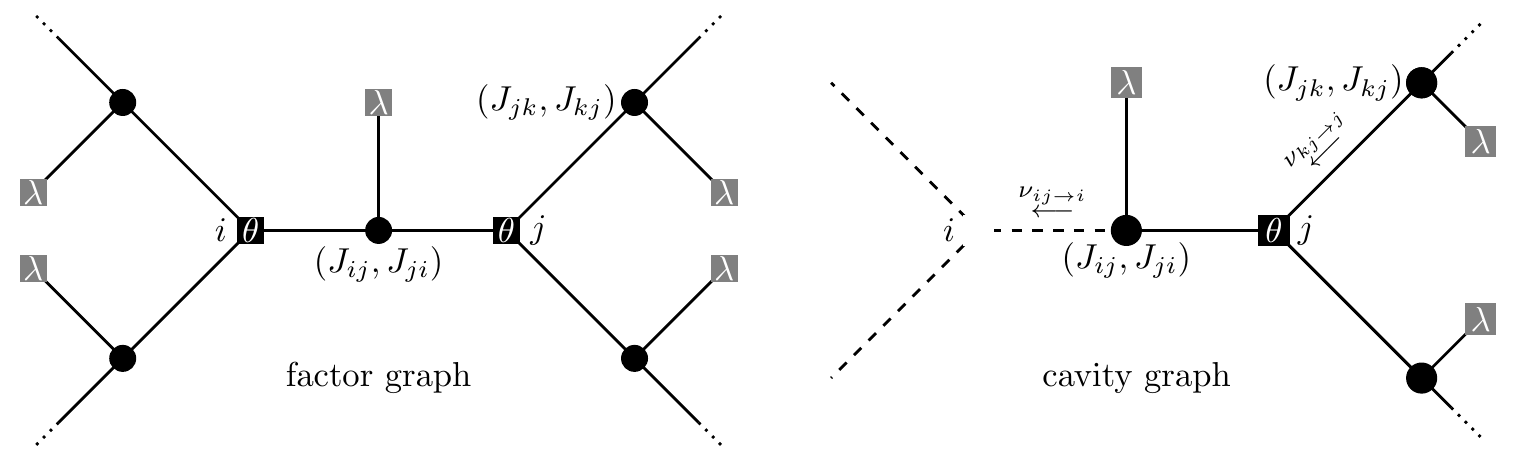}
	\caption{On the left we show the factor graph defining the model for $C=3$. Circular nodes correspond to variable nodes, which are defined by the edges $(i,j)$ of the original graph. There are two type of factor nodes, the light gray one which correspond to the Gaussian distributions in  \eqref{eq:opn-model} correlating $J_{ij}$ and $J_{ji}$ and the black $\theta$ nodes that correspond to the pattern constraints in \eqref{eq:opn-model}. On the right we show the cavity graph where the constraint over node $i$ has been removed.}
	\label{fig:graphs}
\end{figure}

The distribution $\rho(\bJ)$ corresponds to a graphical model where the variable nodes correspond to the edges of the original graph $G$, and the factor nodes correspond to the vertex ones (see Figure \ref{fig:graphs}). We follow a convention and notation similar to matching or assignment problems, (see chapter 16 of \cite{mezard2009information}). 
The observables of interest for this model are the expected symmetry, $\eta$, and the correlation between edge variables, $x$. They are defined by the following formulae,
\begin{align}
\label{eq:eta-def}
    \eta ={}& {}\frac{\expected{J_{ij}J_{ji}}}{\expected{J_{ij}^2}}\,,\\
\label{eq:corr-def}
    x = {}& {} \frac{\expected{J_{ij}J_{ji}} - \expected{J_{ij}}^2}{\expected{J_{ij}^2} - \expected{J_{ij}}^2}\,,
\end{align}
where $\expected{\cdot}$ represents average over the original ensemble (\ref{eq:soft-ensemble}). By symmetry of the model we have assumed $\expected{J_{ij}} = \expected{J_{ji}}$. We can also calculate the free entropy, $\Phi = \frac{1}{N}\log Z$, where $Z$ comes from the normalization of (\ref{eq:opn-model}). In this context $Z$ corresponds to the probability of satisfying the set of hard constraints with a random sample of $\bJ$ distributed according to $\rho_0 (\bJ) = \prod_{(i,j) \in E} \calN_\lambda (\bJ_{ij})$. The formula for $\Phi$ is
\begin{align}
\label{eq:free-entropy-definition}
    \Phi ={} &  \frac{1}{N} \log \int \rmd \bJ \prod_{(i,j)\in E} \calN_{\lambda}(\bJ_{ij})\cdot \prod_{i \in V} \theta(\sum_{j\in\partial i}J_{ij} - \kappa \sqrt{C}).
\end{align}
We see that $Z$ from \eqref{eq:opn-model} is the analog of the fractional synaptic volume \eqref{eq:volume}, therefore the free entropy~\eqref{eq:free-entropy-definition} is the analog of the $\log$ volume calculated with the replica method for the MPN. Once we have established this analogy, we will then consider the free entropy to be a measure of the volume of solutions of the OPN.

We will use the Belief Propagation (BP) formalism for graphical models on tree-like graphs; we refer the reader to \cite{mezard2009information} for an introduction to the subject. The corresponding BP equations for (\ref{eq:opn-model}) are, 
\begin{align}
\label{eq:BPgeneral}
    \nu_{ij \to i}(\bJ_{ij}) \propto {} &\calN_{\lambda} (\bJ_{ij}) \int \prod_{k \in \partial j \setminus i}\rmd\bJ_{jk} \, \theta( J_{ji}  +  \sum_{k\in \partial j\setminus i}J_{jk} - \kappa\sqrt{C}) \prod_{k \in \partial j\setminus i}\nu_{kj \to j}(\bJ_{jk}).
\end{align}

Following the BP formalism, the distribution $
\nu_{ij \to i}(\bJ_{ij})$ corresponds to the joint distribution of the pair $\bJ_{ij} = (J_{ij},J_{ji})$ in a model where the constraint over node $i$ has been removed; we will refer to this distribution as the message from $ij$ to $i$, see Fig.~\ref{fig:graphs}. Since removing a constraint implies removing a factor node from the factor graph defining (\ref{eq:opn-model}), this method is referred to as the \emph{cavity} method. We use the symbol $\propto$ to denote equality up to the normalization of the RHS. In the large $N$ limit we expect these equations to become exact, as loops inducing correlations diverge with the size of the graph like $\Order{\log N}$. The marginals of the original model can then be expressed as
\begin{align}
    \rho_{ij}(\bJ_{ij}) \propto{} & \nu_{ij \to i}(\bJ_{ij})\nu_{ij \to j}(\bJ_{ij})/ \calN_{\lambda}(\bJ_{ij}) .
\end{align}

Many of the observables we will study can be written in terms of this marginal, $\rho_{ij}$. Inspecting equation (\ref{eq:BPgeneral}), one realizes that the whole theory can be written in terms of the marginals of the cavity distribution. We define them as
\begin{equation}
    \begin{aligned}
    \nu_{ij \to i}^L(J_{ij}) ={} & \int \rmd J \, \nu_{ij \to i}(J_{ij},J )\,, \\
    \nu_{ij \to i}^R (J_{ji}) ={} & \int \rmd J \, \nu_{ij \to i}(J ,J_{ji})\,.
\end{aligned}
\end{equation}
The cavity equations can be written as
\begin{equation}
    \begin{aligned}
    \nu^L_{ij \to i}(J_{ij}) ={}& \int \rmd J \cdot \calN(J_{ij}|J) \cdot \nu^R_{ij \to i}(J) \,,\\
    \nu^R_{ij \to i}(J_{ji}) \propto{} &\calN(J_{ji}) \int\prod_{k \in \partial j \setminus i }\rmd J_{jk} \, \theta( J_{ji}  +  \sum_{k\in \partial j\setminus i}J_{jk} - \kappa\sqrt{C}) \prod_{k \in \partial j\setminus i}\nu_{kj \to j}^L(J_{kj})\,,\\
    \nu_{ij\to i}(\bJ_{ij}) ={} & \calN(J_{ij}|J_{ji}) \cdot \nu_{ij \to i}^R(J_{ji})\,,
\end{aligned}
\label{eq:cavity_marginals}
\end{equation}
where $\calN_{\lambda}(J|J')$ corresponds to the conditional distribution of $\calN_{\lambda}(J,J')$ with respect to one of its variables, that is a Gaussian distribution with mean $\lambda J'$ and variance $1-\lambda^2$. The distribution $\calN(J)$ corresponds to the standard Gaussian of zero mean and variance one. The cavity marginals have important and different physical meanings: 
\begin{itemize}
    \item The marginal of the first variable, $\nu_{ij \to i}^L$, corresponds to the distribution of the variable $J_{ij}$ that \emph{is not} constrained by the $\theta$ function; still, it is not completely free, being correlated with $J_{ji}$ that \emph{is} in turn constrained by the $\theta$ function.
    \item  The marginal over the second variable, $\nu_{ij \to i}^R$, corresponds to the distribution of $J_{ji}$, which is the variable that has to satisfy the constraint of the $\theta$ function. Notice that it does not depend directly on the other marginal $\nu_{ij \to i}^L$ on the same node, but only on the ones of the neighbouring nodes.
\end{itemize}
 
 Solving the set of equations~\eqref{eq:cavity_marginals} exactly is quite a challenging task. Even with numerical methods, it requires to simulate the distributions with populations of variables for each message. We simplify our approach by introducing a \emph{Gaussian ansatz} for $\nu_{ij \to i}^L$,
\begin{align}
\label{eq:Gaussian}
    \nu_{ij \to i}^L(J_{ij}) = \calN (J_{ij}|\mu_{ij \to i}, \sigma^2_{ij \to i})\,, 
\end{align}
where the parameters $\mu$ and $\sigma^2$ correspond to the moments of the Gaussian distribution, $\expected{J_{ij}}_{\nu^L} = \mu_{ij \to i}$ and $\textrm{Var}\p{J_{ij}}_{\nu^L} = \sigma^2_{ij \to i}$. 
The accuracy of this approximation, which is reasonable for an unconstrained variable, will be shown \textit{a posteriori}. This also allows to write down an analytic expression for $\nu^R_{ij\to i}$ in terms of known functions, since the sum of Gaussian variables is still Gaussian. Additionally, we assume site symmetry dropping the indices of $\mu$ and $\sigma^2$, since no disorder is present. We can then write down the following set of equations (we define $z = C-1$),
\begin{equation}
    \begin{aligned}
\label{eq:cavity_moments}
    \nu_{ij\to i}^L(J) = {}& \calN(J|\mu,\sigma^2) \,,\\
    \nu_{ij\to i}^R(J) = {}& \frac{\calN(J) H(\kappa \sqrt{C} - z \mu - J)}{\int \rmd J' \calN(J') H(\kappa \sqrt{C} - z \mu - J')} \,, \\
    \mu = {}& \lambda \expected{J}_{\nu^R} \,,\\
    \sigma^2 ={}& 1 - \lambda^2 + \lambda^2 [\expected{J^2}_{\nu^R} - \expected{J}_{\nu^R}^2]\,,
\end{aligned}
\end{equation}
where
\begin{equation}
    H(x) = \frac{1}{2}\erfc\!\left(\frac{x}{\sqrt{2}} \right)\,.
    \label{eq:cavity_H}
\end{equation}

Once we know the messages, we can calculate analytic expressions for many other observables, as for the marginal distribution of the variables on an edge:
\begin{equation}
    \begin{aligned}
    \rho_{ij}(J_{ij},J_{ji}) ={} & \frac{1}{\calZ} H\p{\frac{\kappa\sqrt{C} - z\mu - J_{ij}}{\sqrt{z\sigma^2}}} \calN_{\lambda}(J_{ij},J_{ji}) H\p{\frac{\kappa\sqrt{C} - z\mu - J_{ji}}{\sqrt{z\sigma^2}}}.
\end{aligned}
\end{equation}
With this distribution we can calculate the next fundamental quantities:
\begin{equation}
\label{eq:moments}
    \begin{aligned}
    \expected{J_{ij}^p} ={} &  \int \rmd J \rmd J'   \,J^p\,\rho(J,J') \,,\\
    \expected{J_{ij}J_{ji}} ={} &  \int \rmd J \rmd J'   \,JJ'\,\rho(J,J')\,.
\end{aligned}
\end{equation}

From the previous equations we can then evaluate the asymmetry $\eta$ and the correlation $x$, given by (\ref{eq:eta-def}) and (\ref{eq:corr-def}) respectively. We can also obtain the free entropy~\eqref{eq:free-entropy-definition} from the cavity approach, sometimes in this context referred to as the Bethe free entropy. Following \cite{mezard2009information}, it is given by
\begin{multline}
\label{eq:free-entropy}
    \Phi =  \log H\p{\frac{\kappa\sqrt{C} - C\mu}{\sqrt{C \sigma^2}}} + C\log \int\rmd J   \calN(J) H\p{\frac{\kappa\sqrt{C} - C\mu - J}{\sqrt{z\sigma^2}}} \\
     - \frac{C}{2}\log \int \rmd J \rmd J'  H\p{\frac{\kappa\sqrt{C} - z\mu - J}{\sqrt{z\sigma^2}}} \calN_{\lambda}(J,J') H\p{\frac{\kappa\sqrt{C} - z\mu - J'}{\sqrt{z\sigma^2}}}.
\end{multline}

\begin{figure}
    \centering
    \includegraphics[width=\textwidth]{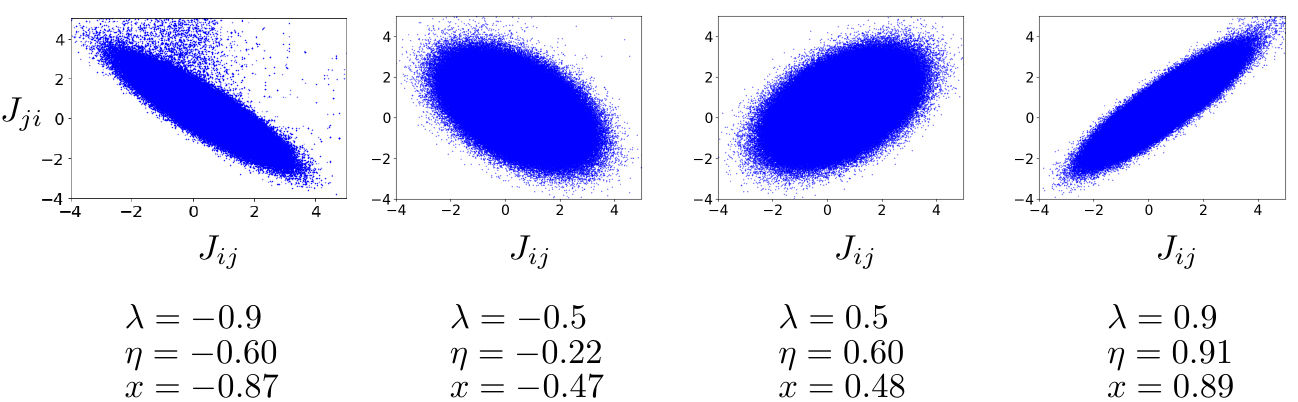}
    \caption{Empirical joint distribution $\rho(J_{ij},J_{ji})$ for different values of $\lambda$ for $N = 1000$, $C =15$, $\kappa = 1.5$. Obtained by MCMC sampling of (\ref{eq:opn-model}), aggregate of 400 samples.} 
    \label{fig:samples}
\end{figure}

\begin{figure}
    \centering
    \includegraphics[width=\textwidth]{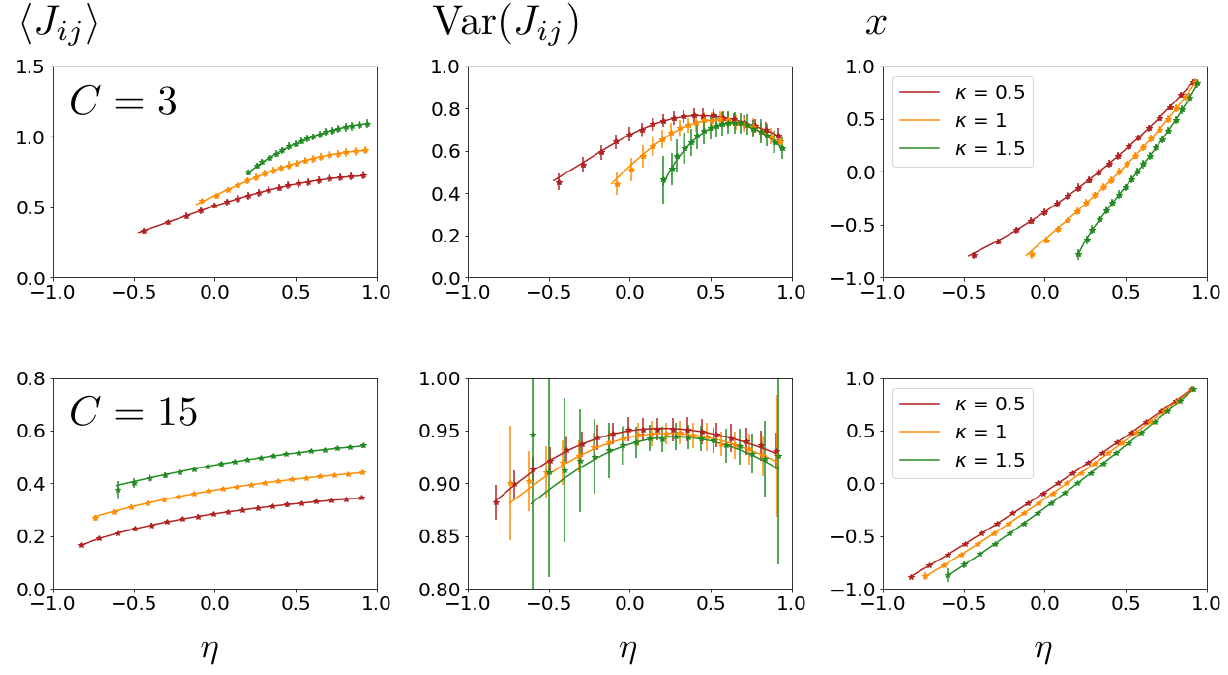}
    \caption{We compare the moments of the marginal distribution $\rho(J_{ij},J_{ji})$ as measured in MCMC simulations (symbols) against the result of the numerical solution (solid lines) of the cavity equations \eqref{eq:cavity_moments}  applied to \eqref{eq:moments}. Top row corresponds to a model with a graph of connectivity $C = 3$ and bottom row to connectivity $C = 15$. {Averages were made over 400 networks. Error bars are present in all plots even when not visible.}}
    \label{fig:moments-MCMCvsSimulation}
\end{figure}

In order to compare our theory with the original model (\ref{eq:opn-model}), we sampled from it numerically using standard Markov Chain Montecarlo (MCMC) techinques {(see Appendix~\ref{sec:numerics} for details)}. The set of equations for the cavity moments (\ref{eq:cavity_moments}) can be solved numerically by iteration with a learning rate. In Fig.~\ref{fig:samples}  we plot the joint distribution of pairs $(J_{ij},J_{ji})$ obtained numerically. The empirical moments obtained from the samples are compared to the ones predicted by the theory, (\ref{eq:moments}), in Fig.~\ref{fig:moments-MCMCvsSimulation}. Plots are made against $\eta$, evaluated from (\ref{eq:eta-def}), as the asymmetry is the physically relevant parameter.  We observe an excellent agreement between theory and simulation, proving that the Gaussian approximation \eqref{eq:Gaussian} was a reasonable choice.

We observe the existence of a critical value of asymmetry, $\eta_c$, below which (\ref{eq:opn-model}) is not able to sample solutions of the set of constraints. This critical value is achieved when $\lambda = -1$, corresponding to the point where the distribution $\calN_{\lambda}(J,J')$ becomes singular by concentrating in a line. If we look at the correlation, we see that indeed it is tending to $x = -1$ as well. Nevertheless, in general we have $\expected{J_{ij}} > 0$ up to the critical value of $\eta$. This means that the distribution of synaptic weights is not centered at zero, and in the critical case of $x = -1$ it concentrates on a straight line not going through the origin. This explains the difference between the asymmetry $\eta$ and the correlation $x$: if the joint distribution of $(J_{ij},J_{ji})$ is not zero-mean, then we can have $x = -1$ while $-1<\eta$. This effect can be observed in Figure \ref{fig:samples} as $\lambda$ approaches $-1$. Notice that the opposite limit of perfect correlation, $x = 1$, \emph{can always} reach $\eta = 1$, since a finite mean is not incompatible with this condition.

\begin{figure}
    \centering
\includegraphics[width=.95\textwidth]{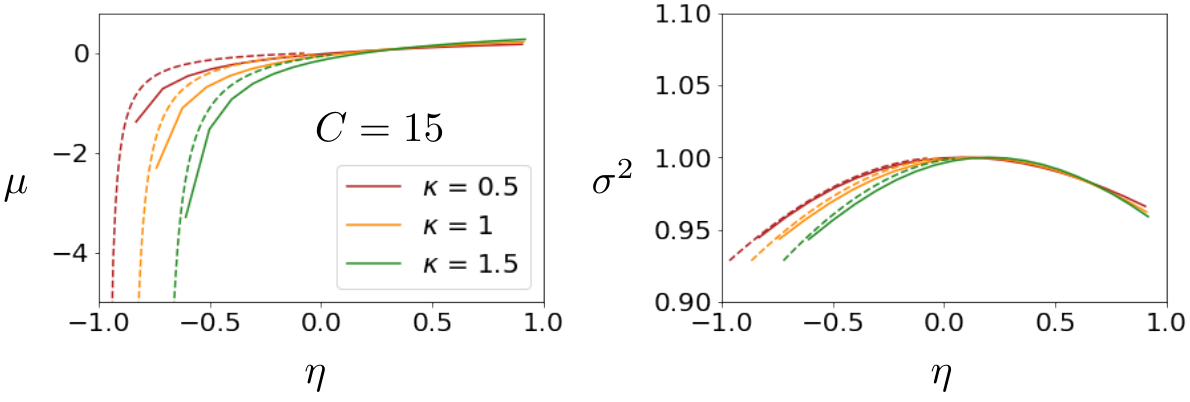}
    \caption{We show the numerical solutions for $\mu$ and $\sigma^2$ (\ref{eq:cavity_moments}) for $C = 15$ in solid lines. Dashed lines correspond to the solutions of the asymptotic equations \eqref{eq:asymptotic-equations}.}
    \label{fig:cavity-moments}
\end{figure}

Interestingly, the behaviour of the model at $x=1$ is completely different to that at $x=-1$. When the model reaches perfect correlation $x=1$, it cannot go beyond $\eta=1$ simply because this is the maximum value for $\eta$. Both moments $\mu$ and $\sigma^2$ remain finite, and consequently the same for the volume of solutions, \eqref{eq:free-entropy}. For the case of perfect anti-correlation $x=-1$, when $\eta$ approaches $\eta_c>-1$ we observe signals of a SAT/UNSAT transition: the volume of solution vanishes and all the constraints become marginally satisfied. This can be seen by looking at the behaviour of $\mu$ and $\sigma^2$.  When solving \eqref{eq:cavity_moments} for values of $\lambda$ close to $-1$, we observe that $\mu$ diverges to $-\infty$ while $\sigma^2$ remains finite as $\eta\to\eta_c$ (see Fig.~\ref{fig:cavity-moments}). This means that at any point before the critical point the width of the cavity distribution remains finite and is not shrinking, while its mean is diverging. In Sec.~\ref{subsec:asympt_lambda} we will show that this divergence exists within our theory doing an asymptotic analysis close to $\lambda = -1$. The divergence of $\mu \to -\infty$ implies that at the critical point $\eta_c$ the volume of solutions vanishes, since the first term in \eqref{eq:free-entropy} diverges to $-\infty$ while the other terms cancel each other,
\begin{align}
    \lim_{\mu\to -\infty} \Phi = \lim_{\mu \to -\infty } \log H\!\p{\frac{\kappa\sqrt{C} - C\mu}{\sqrt{C \sigma^2}}} = -\infty .
\end{align}

\begin{figure}
    \centering
    \includegraphics[scale=1]{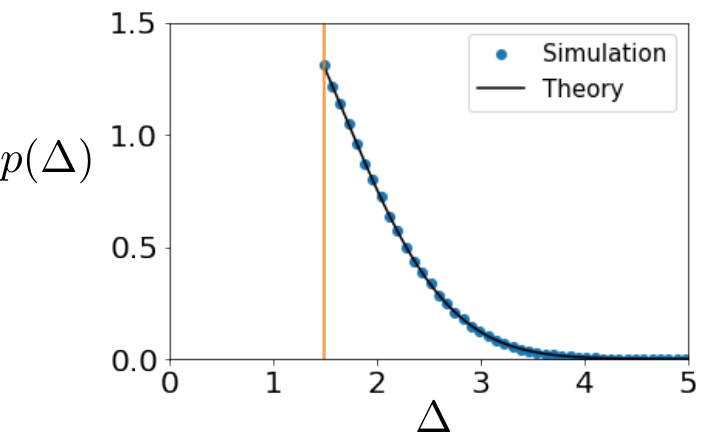}
    \caption{Distribution of gap variable $\Delta_i$, (\ref{eq:gaps}). The blue dots correspond to the histogram obtained from MCMC on a model with $N = 1000$, $C = 15$, $\kappa = 1.5$, $\eta = 0.6$. The solid black line corresponds to theoretical estimate (\ref{eq:gap-distribution}). The orange vertical line is at $\Delta = \kappa$. Aggregate of 400 samples ($4\times 10^5$ values of $\Delta_i$).}
    \label{fig:gap-distribution}
\end{figure}

To see that all constraints become marginally satisfied at $\eta_c$, meaning they all turn into equalities, we look at the gap variables, defined as
\begin{align}
\label{eq:gaps}
    \Delta_i = \frac{1}{\sqrt{C}}\sum_{j \in \partial i} J_{ij}.
\end{align}
We calculate the distribution of gaps from the joint distribution of variables around a given factor node $i$, $\bJ_{\partial i} = (\bJ_{ij})_{j\in\partial i}$, which is given by the product of the messages truncated by the constraint in $i$,
\begin{equation}
    \begin{aligned}
    p(\Delta) \propto{} & \int \prod_{j\in\partial i}\rmd \bJ_{ij}\, \delta(\Delta - \frac{1}{\sqrt{C}}\sum_{j\in\partial i} J_{ij})\, \theta(\sum_{j\in\partial i} J_{ij} - \kappa\sqrt{C}) \prod_{j\in \partial i} \nu_{ij \to i} (\bJ_{ij}) \,,\\
    \propto{} & \theta(\Delta\sqrt{C} - \kappa\sqrt{C}) \int \prod_{j\in \partial i}\rmd J_{ij}\, \delta(\Delta - \frac{1}{\sqrt{C}}\sum_{j\in\partial i} J_{ij}) \prod_{j\in\partial i}\calN(J_{ij} | \mu, \sigma^2) \,,\\
    \propto{} & \theta(\Delta - \kappa)\, \calN(\Delta | \sqrt{C}\mu, \sigma^2)\,.
\end{aligned}
\end{equation}
This means that $p(\Delta)$ corresponds to a Gaussian truncated at the margin $\kappa$.
\begin{align}
\label{eq:gap-distribution}
    p(\Delta) = {}& \frac{\theta(\Delta - \kappa) \calN(\Delta | \sqrt{C}\mu, \sigma^2)}{H\!\p{\p{\kappa - \sqrt{C}\mu}/{\sigma}}}\,.
\end{align}
This expression can be compared with the empirical distribution of gaps obtained from numerical sampling. In Fig.~\ref{fig:gap-distribution} we can see there is very good agreement. In the next section we will show how this distribution becomes a $\delta$-function peaked at $\kappa$ when $\eta \to \eta_c$. This result shows how the mean, $\mu$, and variance, $\sigma^2$, of the cavity distribution $\nu_{ij \to i}$ can be actually measured numerically from samples of (\ref{eq:opn-model}), since they are related to the moments of the gap distribution by
\begin{equation}
    \begin{aligned}
    \expected{\Delta} = {}&\sqrt{C} \mu + \sigma \frac{\calN(\kappa|\sqrt{C}\mu, \sigma^2)}{H\!\p{\p{\kappa - \sqrt{C}\mu}/{\sigma}}}, \\
    \textrm{Var}(\Delta) = {}& \sigma^2\left[ 1 +  \frac{(\kappa - \sqrt{C}\mu)\calN(\kappa|\sqrt{C}\mu, \sigma^2)}{\sigma H\!\p{\p{\kappa - \sqrt{C}\mu}/{\sigma}}} - \p{\frac{\calN(\kappa|\sqrt{C}\mu, \sigma^2)}{H\!\p{\p{\kappa - \sqrt{C}\mu}/{\sigma}}}}^2 \right]\,.
\end{aligned}
\end{equation}

\subsection{Asymptotic behavior}
\label{subsec:asympt_lambda}

In order to get a better understanding of the theory, we look at asymptotic expressions close to $\eta_c$. For this, it is necessary to assume $\lambda$ very close to $-1$, $0<\lambda+1 \ll 1$. Performing a Laplace approximation we can get an approximate form of the equations,
\begin{equation}
    \begin{aligned}
\label{eq:asymptotic-equations}
    \mu = {}& \lambda \frac{\kappa\sqrt{C} - z \mu}{1 + z \sigma^2},\\
    \sigma^2 = {}& 1 - \lambda^2 + \lambda^2 \p{\frac{1}{1 + \frac{1}{z\sigma^2}}}.
\end{aligned}
\end{equation}
Looking at the leading order in $1/(1+\lambda)$ we get a divergence for $\mu$ at $\lambda = -1$ and a finite value for $\sigma^2$, as observed in simulation: 
\begin{equation}
    \begin{aligned}
\label{eq:asymptotic-values}
    \mu = {}& \frac{\lambda \kappa\sqrt{C}}{(C-1)( 1 + \lambda)}, \\
    \sigma^2 = {}& \frac{C-2}{C-1}.
\end{aligned}
\end{equation}

Doing similar approximations for the moments in (\ref{eq:moments}) and using the solutions of (\ref{eq:asymptotic-equations}), we can then deduce the critical values at $\lambda = -1$,
\begin{equation}
    \begin{aligned}
\label{eq:asymptotic-moments}
    \eta_c = {}& \frac{\kappa^2  -C +2 }{\kappa^2 + C - 2} \,,\\
    x_c = {}& -1 \,,\\
    \expected{J_{ij}}_c = {}& \frac{\kappa}{\sqrt{C}} \,,\\
    \textrm{Var}\p{J_{ij}}_c = {}&  \frac{C-2}{C}.
\end{aligned}
\end{equation}
This confirms what is happening at $\lambda = -1$: the synaptic variables become completely anti-correlated; nevertheless, they retain a finite value for the mean and the variance. This means that the entries should obey the following relationship, 
\begin{align}
    J_{ij} - \expected{J_{ij}}_c =  - J_{ji} + \expected{J_{ij}}_c.
\end{align}

Additionally, as we mentioned before, at criticality all row constraints turn into equalities. If we assume that $\mu\to - \infty$, it can be shown through manipulations of the gap distribution that the latter is becoming a $\delta$-function centered at $\kappa$,
\begin{align}
    p(\Delta) \approx \theta(\Delta - \kappa) \rme^{-\sqrt{C}\frac{|\mu|}{\sigma^2}(\Delta - \kappa)}\frac{\sqrt{C}|\mu|}{\sigma^2}\underset{\mu\to -\infty}{\longrightarrow} \delta(\Delta - \kappa).
\end{align}
The picture then is that as $\lambda$ approaches $-1$, the measure (\ref{eq:opn-model}) starts concentrating along the set of solutions to the system of equations
\begin{equation}
    \begin{aligned}
\label{eq:marginal-equations}
    \sum_{j\in\partial i} J_{ij} = {}& \kappa \sqrt{C} \,,\\
    J_{ji} = {}&  -J_{ij} + 2 \frac{\kappa}{\sqrt{C}} \,.
\end{aligned}
\end{equation}
We show that this system admits solutions by decomposing the matrix in its symmetric, $J_{ij}^{s} = (J_{ij} + J_{ji})/2$  and antisymmetric part, $J_{ij}^{a} = (J_{ij} - J_{ji})/2$. The symmetric part is constant, $J^s_{ij} = \kappa/\sqrt{C}$ and the antisymmetric part satisfies the set of equations,
\begin{align}
\label{eq:antisymmetric-equation}
    \sum_{j\in\partial i} J^a_{ij} = 0  \qquad \forall i\,.
\end{align}

Solutions for these equations do exist: they correspond to flows on the graph without sinks or sources. Even though there are possibly infinitely many solutions, clearly all the $J_{ij}$'s are highly correlated due to the strong constraints of \eqref{eq:antisymmetric-equation} around each node. Therefore, as $\lambda$ approaches $-1$ the volume vanishes because the effective number of degrees of freedom becomes smaller than $NC/2$, which corresponds to the number of degrees of freedom imposed only by the antisymmetry. Nevertheless, since this volume is concentrating in a continuous set of solutions, we still have a finite variance at the critical point. 

Notice that other asymptotic limits can be studied, not only the case of $\mu\to-\infty$ as $\lambda\to -1$. One can also look at the cases $\kappa\to\infty$ and $C \to\infty$. We first explore the $\kappa \gg 1$ case, which uses the same asymptotic expansion of $H(x)$, so we can use equations (\ref{eq:asymptotic-equations}). In this case, we see that even the most anti-correlated solutions, $x\approx -1$, posses a very high degree of symmetry, $\eta \approx 1$, since the critical asymmetry $\eta_c$ approaches $1$ very quickly,
\begin{align}
    \eta_c \approx 1 - 2 \frac{C-2}{\kappa^2 }\,.
\end{align}
One can also see that the distribution of $J_{ij}$'s is concentrating in its mean, which is actually independent of $\lambda$, 
\begin{align}
\label{eq:mean-large-kappa}
    \expected{J_{ij}} = \frac{\kappa}{\sqrt{C}}\,,
\end{align}
because the variance-to-mean ratio is vanishing with $\kappa$, 
\begin{align}
    \frac{\Var(J_{ij})}{\expected{J_{ij}}} \sim \frac{1}{\kappa} \underset{\kappa\to \infty}{\longrightarrow} 0 \,.
\end{align}
This means that for all values of correlation, all the solutions are very close to the simple solution, $J_{ij}\approx\kappa/\sqrt{C}$. The effect of this can also be seen in the gap distribution that is going to $p(\Delta) = \delta(\Delta - \kappa)$ as $\kappa\to\infty$ as well. 

Finally, we will explore the $C \to \infty$ limit. This will actually allow us to directly connect our results with those of GGY. First, it is important to notice that increasing $C$ actually makes the problem \emph{easier}. This is due to the fact that when $C$ grows the number of variables, $NC$, is increasing while keeping fixed the number of hard constraints, $N$. The exact ratio of constraints over variables is exactly $\alpha = 1/C$. When taking the limit $C\to \infty$ in equations (\ref{eq:cavity_moments}) and (\ref{eq:moments}), all integrals can be performed exactly by the Laplace method, giving
\begin{equation}
    \begin{aligned}
    \eta = {}& \lambda \,,\\
    \mu = {}& \expected{J_{ij}} = 0 \,,\\
    \sigma^2 = {}& \expected{J^2_{ij}} = 1 \,.
\end{aligned}
\end{equation}

This means that there is no apparent effect of the constraints at the level of the marginal $\rho_{ij}(J_{ij},J_{ji})$, and all asymmetries can be sampled, $\eta\in(-1,1)$. Nevertheless there is an effect, and this can be seen by looking at the distribution of gaps, (\ref{eq:gap-distribution}), $p(\Delta)$, which is controlled by the product $\sqrt{C} \mu$. If we define this parameter as $r = \sqrt{C}\mu$, it is actually natural to assume that it will be of $\Order{1}$ in the $C\to \infty$ limit, just by looking, for example, at (\ref{eq:asymptotic-values}). With the Laplace method we can derive an exact equation for $r$ multiplying (\ref{eq:asymptotic-equations}) by $\sqrt{C}$ and taking the limit $C\to\infty$,
\begin{align}
\label{eq:r-equation-C-large0}
    r = \lim_{C\to\infty} \sqrt{C}\lambda\frac{\int \rmd J_{0}\,J_{0}\,\calN(J_{0})H\p{\p{\kappa\sqrt{C} - r\sqrt{C} - J_{0}}/{\sqrt{z\sigma^2}}}}{\int \rmd J_{0}\,\calN(J_{0})H\p{\p{\kappa\sqrt{C} - r\sqrt{C} - J_{0}}/{\sqrt{z\sigma^2}}}}.
\end{align}
Knowing that $\lambda\to\eta$ and $\sigma^2\to 1$, we find the following equation for $r$,
\begin{align}
\label{eq:r-equation-C-large}
    r = \eta\frac{\rme^{-\frac{1}{2}(\kappa -r)^2}}{\int_{\kappa - r}^\infty \rmd t\,\rme^{-\frac{1}{2}t^2}}
\end{align}
Even though we derived this equation starting from a Gaussian approximation of the message $\nu_{ij \to i}$, \eqref{eq:Gaussian}, this result for $C\to \infty$ is actually exact. It can be derived directly through a replica calculation for the OPN, as shown in Appendix \ref{app:opn-replica}. Once $r$ is determined, the gap distribution can be calculated in the following way:
\begin{align}
\label{eq:gap-distribution-r}
    p(\Delta) = {}& \frac{\theta(\Delta - \kappa) \rme^{-\frac{1}{2}(\Delta - r)^2}}{\int_{\kappa}^\infty\rmd t \, \rme^{-\frac{1}{2}(t - r)^2}}\,.
\end{align}
The free entropy gets the simpler form,
\begin{align}
	\Phi = & \log H(\kappa - r) - \frac{r^2}{2\eta}\,.
\end{align}

The asymptotic solution for $r$  close to $\eta = -1$ is  $r \approx \kappa \eta / ( 1+ \eta)$. This means that still there is no solution for $r$ at $\eta = -1$, therefore a critical point is present also in the $C\to \infty$ case. The volume of solutions is still vanishing, as the free entropy is still diverging to $-\infty$. This is noteworthy since it means that even storing one pattern with $\kappa>0$ and perfect anti-symmetry is impossible for $C\to\infty$, which is the same as the limit $\alpha\to0$. Even though the gap distribution is going to $p(\Delta) = \delta(\Delta - \kappa)$ when $\eta \to -1$, this does not mean that for large enough $C$ there exist solutions for the systems of equations $\sum_{j\in\partial i}J_{ij} =\kappa\sqrt{C}$ with $\eta = -1$, that is $J_{ij} = -J_{ij}$. As a matter of fact, it can be shown that there are no solutions except for $\kappa = 0$, as discussed in \cite{gardner1989phase}. If one assumes a large but finite $C$, the asymptotic expansion used to derive (\ref{eq:r-equation-C-large}) is only valid as long as $r\ll\sqrt{C}$. As soon as we assume a finite value of $C$, a value $\kappa>0$ will shift the critical value to $\eta_c>-1$, and the discussion stemming from (\ref{eq:marginal-equations}) and (\ref{eq:antisymmetric-equation}) applies. The relationship between $\eta_c$ and $\alpha = 1/C$,  (\ref{eq:asymptotic-moments}), can be expanded around $\alpha = 0$, which gives
\begin{align}
\label{eq:critical-line}
    \eta_c = - 1 + 2\kappa^2 \alpha.
\end{align}
Amazingly, this is exactly the same result obtained by GGY  in \cite{gardner1989phase} for the MPN, for $\alpha$ defined as the ratio of number of constraints over number of variables. We will explore this connection in the next sections.

\section{Many patterns network: replica approach \label{sec:replica}}

In the last section we discussed the existence of a non-trivial critical value of the degree of asymmetry $\eta$ where the OPN constraint satisfaction problem with soft asymmetric constraints ceases to have solutions. In this section we will go back to the problem with many patterns $P = \alpha C$ and hard constraints, originally discussed in~\cite{gardner1989phase}, which we call many patterns network (MPN). With the replica analysis we will report in this section and in analogy with the OPN case, we will argue that the peculiar nature of the satisfiability transition for very asymmetric synaptic matrices is not related to a clustering transition occurring in the space of solutions of the problem, as suggested by GGY. The replica analysis is particularly suited to obtain this kind of information for a typical instance a random CSP: the occurrence of replica symmetry breaking in this context can be explained as a clustering transition in the space of solutions, see~\cite{krzakala2007}.

 The main original contributions in this section are reported in \ref{subsec:fRSB}, where a full-replica-symmetry-breaking (fullRSB) analysis of the problem is devised and the saddle-point equations~\eqref{eq:saddle_r}--\eqref{eq:saddle_h} for the variational order parameters (the overlap functions between solutions and the auxiliary field $r$ introduced in~\eqref{eq:replica_freeEnergy0}) can be found, together with equation~\eqref{eq:fullRSB_replicon} stating the \emph{marginal stability} of the fullRSB solution under variations in the replica space. These results will be applied in~\ref{subsec:RS} to push forward the replica-symmetric analysis by GGY and better characterize the phase space of the problem. 

To make the discussion more self-contained, we start here reviewing the GGY approach to this CSP in the replica framework, which will result in an expression for the free energy of the replicated system for arbitrary overlap matrices, Eq.~\eqref{eq:S_replica}. The quenched average over the pattern distribution of the logarithm of the volume~\eqref{eq:volume} is performed via the identity
\begin{equation}
    \mean{\log V} = \lim_{n\to 0} \frac{\mean{V^n} - 1}{n}\,.
    \label{eq:replica_trick}
\end{equation}
We use the overline notation for the average over the random memory patterns $\xi^\mu$. Once the integer moments $\mean{V^n}$ are obtained for $n\in \mathbb{N}^+$, a continuation prescription for $n\to 0$ must be chosen carefully. 

However, as noted by GGY, the application of the replica method to the present model is not straightforward and requires to make a crucial assumption, the hypothesis of site symmetry, whose validity we will postulate in this section. Indeed, due to the definite-asymmetry constraint~\eqref{eq:symm_constr}, which introduces correlations between the couplings $J_{ij}$ and $J_{ji}$, the contributions from different ``sites'' $i$, $j$ do not factorize after the average over the disorder, and the resulting replicated volume cannot be written only in terms of global overlaps between replicas. In order for this to be the case, not only the overlap matrices
\begin{equation}
    q_{ab}^{i} = \frac{1}{C} \sum_{j\in \partial i} J_{ij}^a J_{ij}^b\,, \qquad h_{ab}^{i} = \frac{1}{C} \sum_{j\in \partial i} J_{ij}^a J_{ji}^b\,,
    \label{eq:overlaps}
\end{equation}
where $a$ and $b$ are replica indices, must not depend on the index $i$, but the product itself $J_{ij}^a J_{ji}^b$ need to be substituted with its average over the sites,
\begin{equation}
\label{eq:mean-field-assumption}
    J_{ij}^a J_{ji}^b \to \frac{1}{N} \sum_{k}J_{kj}^a J_{jk}^b\,,
\end{equation}
in a mean-field fashion. In this section, we will assume the validity of this site-symmetric assumption and we will focus on the stability of the replica symmetric solution of the resulting model, postponing a partial re-examination of this point to Sec.~\ref{sec:discussion}.

With standard manipulations (see App.~\ref{app:replica}) we can write the replicated volume as
\begin{equation}
    \mean{V^n}= \int \prod_{a<b} \diff q_{ab}\, \diff h_{ab} \,\rme^{nCNS[\hat{q},\hat{h}]}\,,
\end{equation}
where the $n\times n$ matrices $\hat{q}$, $\hat{h}$ are introduced via Eq.~\eqref{eq:overlaps} and are such that $q_{aa} = 1$ (due to the spherical constraints over the rows of the synaptic matrix), $h_{aa} = \eta$ (due to the definite-asymmetry constraints); {$S$ is a function of the overlap matrices $\hat{q}$, $\hat{h}$ whose value at the saddle point gives the typical value of the free energy density of the $n$-times replicated problem: in the limit $n\to 0$, it gives back the typical free energy of the original model with quenched disorder, according to Eq.~\eqref{eq:replica_trick}. The form of $S$, to be determined variationally with respect to its arguments,} is given by
\begin{multline}
\label{eq:replica_freeEnergy0}
    nCNS[\hat{q},\hat{h}] = \frac{CN}{4} \log \det(\hat{q}^2 - \hat{h}^2) - \frac{nCN}{4} \log (1 - \eta^2)\\
    + \alpha C \log \int \prod_{a}\diff r^a\, \rme^{- \frac{N}{2}\sum_{a,b}h^{-1}_{ab} r^a r^b+ N \log\bigr[\rme^{\frac{1}{2}\sum_{a,b}q_{ab} \frac{\partial^2}{\partial y^a \partial y^b} }\prod_{a}\theta(y^a ) \bigr|_{y^a=r^a - \kappa}\bigr]}\,,
\end{multline}
The auxiliary variables $r^a$ are introduced in order to completely decouple the sites $i$, $j$ in the calculation, via a Hubbard-Stratonovich transformation: we will see in the following how these quantities are related to the rescaled average of the cavity distribution, $r=\sqrt{C}\mu$, appearing in Eq.~\eqref{eq:r-equation-C-large0} for the OPN. The inner integrals in $r^a$ can be solved via a saddle-point method for large $N$, which is located at
\begin{equation}
    \sum_{b} h^{-1}_{ab}r_{b} = \partial_{r^a}\log\bigr[\rme^{\frac{1}{2}\sum_{a,b}q_{ab} \frac{\partial^2}{\partial y^a \partial y^b} }\prod_{a}\theta(y^a ) \bigr|_{y^a=r^a - \kappa}\bigr]\,.
\end{equation}
This means that we can write the replicated Gardner volume as
\begin{equation}
    \mean{V^n}= \int \prod_{a<b} \diff q_{ab}\, \diff h_{ab} \int \prod_a \diff r^a \!\diff \tilde{r}^a \,\rme^{nCNS[\hat{q},\hat{h},r,\tilde{r}]}\,,
\end{equation}
where now the free energy is given by
\begin{equation}
\begin{aligned}
    nS[\hat{q},\hat{h},r,\tilde{r}] ={}&{}  \frac{1}{4} \log \det(\hat{q}^2 - \hat{h}^2) - \frac{n}{4} \log(1 - \eta^2)  - \alpha \frac{1}{2}\sum_{a,b}h^{-1}_{ab} r^a r^b \\
    {}&{}+ \alpha \log\bigr[\rme^{\frac{1}{2}\sum_{a,b}q_{ab} \frac{\partial^2}{\partial y^a \partial y^b} }\prod_{a}\theta(y^a ) \bigr|_{y^a=r^a - \kappa}\bigr] +\alpha \sum_{a,b} \tilde{r}^a h^{-1}_{ab} r^b\\
    {}&{}- \alpha \sum_{a}\tilde{r}^a \partial_{r^a}\log\bigr[\rme^{\frac{1}{2}\sum_{a,b}q_{ab} \frac{\partial^2}{\partial y^a \partial y^b} }\prod_{a}\theta(y^a) \bigr|_{y^a=r^a - \kappa}\bigr]\,.
\end{aligned}
\end{equation}
The variables $\tilde{r}^a$ have the role of Lagrange multipliers enforcing the saddle-point value of $r^a$. However, at the saddle-point $\tilde{r}^a = 0$, so we can neglect these contributions and consider $S$ as a function of the matrices $\hat{q}$, $\hat{h}$ and of the vector $r^a$ in replica space.

In the following, we will always assume replica symmetry on the vector sector of the model, $r^a = r$ for any $a$, an hypothesis motivated by the hope that thermodynamic observables (which are vectors in replica space) are self-averaging. Then the variational free energy becomes
\begin{equation}
\begin{aligned}
    nS[\hat{q},\hat{h},r] ={}&{} \frac{1}{4} \log \det(\hat{q}^2 - \hat{h}^2) -\frac{n}{4} \log (1 - \eta^2)  - \alpha \frac{r^2}{2}\sum_{a,b} h^{-1}_{ab} \\
    {}&{}+ \alpha \log\bigr[\rme^{\frac{1}{2}\sum_{a,b}q_{ab} \frac{\partial^2}{\partial y^a \partial y^b} }\prod_{a}\theta(y^a ) \bigr|_{y^a=r - \kappa}\bigr]\,.
\end{aligned}
\label{eq:S_replica}
\end{equation}

To proceed further and take the $n\to 0$ limit in Eq.~\eqref{eq:replica_trick}, one should impose an ansatz on the form of the matrices $\hat{q}$, $\hat{h}$ in the replica space. The solution of the problem under the replica-symmetric anstaz (RS)
\begin{equation}
    q_{ab} = (1-q_0)\delta_{ab} + q_0\,, \qquad h_{ab} = (\eta-h_0)\delta_{ab} + h_0
\end{equation}
is studied in great detail by GGY. As in other constraint satisfaction problems, a critical curve $\alpha = \alpha_c(\eta,\kappa)$ (storage capacity) can be defined as the line where a SAT/UNSAT transition occurs, that is where $S$ goes to $-\infty$ and the volume of solutions shrinks to 0. 

This transition presents however unusual features with respect to the case with no correlation between different couplings, studied in the seminal paper~\cite{gardner1987perceptron}. In that case, indeed, the transition always takes place for $q_0 \to 1$: whenever the number of solutions of the CSP decreases, different replicas of the system become more and more correlated and their overlap tends to 1, at the point where a single solution is left. This picture remains unchanged even in CSPs where the RS ansatz is not correct (see for example~\cite{franz2017,franz2019multi,GEPR2020PRE,GPR2020PRL}), where the RS $q_0 \to 1$ line gives an upper bound on the true $\alpha_c$ where the transition occurs.

In the present case, instead, at fixed $\kappa$ there is an interval of values for $\eta$ where the satisfiability transition, that is the divergence of the free energy~\eqref{eq:S_replica} corresponding to the disappearance of solutions, occurs at values of the overlap $q_0$ strictly less than 1. This phenomenon is mathematically due to the fact that the auxiliary variable $r$ is going to $-\infty$ at a finite value of $(1-q_0)$. 

This unnatural result is interpreted by GGY as a sign of replica symmetry breaking: the volume should shatter into disconnected regions and, in the limit of capacity, there should be two, or more, isolated solutions with a non-trivial overlap. We dedicate the rest of this section to the exploration of this possibility. To do so, in the following we will write the free energy~\eqref{eq:S_replica} in a full-replica-symmetry-breaking (fullRSB) scheme. The RS solution of~\cite{gardner1989phase} will be obtained in Sec.~\ref{subsec:RS} directly from the fullRSB one: this will give us the opportunity to review the known results on the RS phase diagram we mentioned above. Moreover, demanding consistency of the fullRSB equations around the RS solution, we will be able to study the stability of the RS ansatz, pushing forward the analysis of GGY and motivating our claim that replica symmetry breaking, though present in a certain region of the parameters space, is not related to the unusual SAT/UNSAT transition occurring at $q_0<1$.

\subsection{Variational fullRSB free energy}
\label{subsec:fRSB}

Following Parisi's fullRSB scheme~\cite{parisi1980}, replica symmetry is broken in $K$ steps with a hierarchy of ans\"atze on the form of the matrices $\hat{q}$, $\hat{h}$. When these steps becomes infinitely many, both the overlap matrices are parametrized formally by their diagonal elements and by monotonic functions depending on a variable $t \in [0,1]$. We can write $\hat{q}\to (1,q(t))$, $\hat{h}\to(\eta,h(t))$, with
\begin{equation}
    q(t) = \begin{cases}
    q_m & \text{if $0<t<t_m$}\\
    q(t) & \text{if $t_m<t<t_M$}\\
    q_M & \text{if $t_M<t<1$}
    \end{cases}\,,\qquad
    h(t) = \begin{cases}
    h_m & \text{if $0<t<t_m$}\\
    h(t) & \text{if $t_m<t<t_M$}\\
    h_M & \text{if $t_M<t<1$}
    \end{cases}\,.
\end{equation}
To write the free energy~\eqref{eq:S_replica} in the fullRSB scheme, it is useful to introduce the quantities
\begin{equation}
    \Lambda_\pm(t) = 1\pm\eta - t[q(t) \pm h(t)] - \int_t^1 \diff u [q(u) \pm h(u)]\,,
\end{equation}
which are sometimes called in literature ``replica Fourier transforms''~\cite{dedominicis1997}, in this case of the matrices $\hat{q}\pm \hat{h}$ whose product appears in the determinant in Eq.~\eqref{eq:S_replica}.
Following~\cite{sommers1984variational} we can write, in the limit of the number of replicas $n\to 0$, the free energy as a functional of $q(t)$, $h(t)$, that is
\begin{equation}
\begin{aligned}
    S[q(t),h(t),r] = {}&{} \frac{1}{4} \log[\Lambda_+(1)\Lambda_-(1)] + \frac{(q_m+h_m)\Lambda_-(0)+(q_m-h_m)\Lambda_+(0)}{4 \Lambda_+(0) \Lambda_-(0)}\\
    {}&{}+  \int_0^1 \diff t \frac{[\dot{q}(t)+\dot{h}(t)]\Lambda_-(t) + [\dot{q}(t)-\dot{h}(t)]\Lambda_-(t)}{4\Lambda_+(t)\Lambda_-(t) }\\
    {}&{} - \frac{\alpha r^2}{2} \left[\eta - \int_0^1 \diff t \,h(t)\right]^{-1} 
    + \alpha \gamma_{q_m} \star f(0,y) \bigr|_{y=r - \kappa} \\
    {}&{}- \alpha \int \diff y\, P(1,y) \left[f(1,y) - \log \gamma_{1-q_M} \star \theta(y) \right]\\
    {}&{}+ \alpha \int \diff y \int_{t_m}^{t_M} \diff t \, P(t,y) \left\{\dot{f}(t,y) +\frac{1}{2} \dot{q}(t) \left[f''(t,y) + t f'(t,y)^2\right] \right\} \,,
\end{aligned}
\label{eq:S_func}
\end{equation}
where the Gaussian convolution is defined as
\begin{equation}
    \gamma_q \star g(y) = \rme^{\frac{q}{2}\frac{\diff^2}{\diff y^2}} g(y) = \int \frac{\diff \nu}{\sqrt{2\pi q}} \rme^{-\frac{\nu}{2q}} g(y-\nu)\,.
\end{equation}
The function $f$ is obtained from the term
\begin{equation}
    \frac{1}{n}\log\bigr[\rme^{\frac{1}{2}\sum_{a,b}q_{ab} \frac{\partial^2}{\partial  y^a \partial y^b} }\prod_{a}\theta(y^a ) \bigr|_{y^a=r - \kappa}\bigr] \,\,\underset{n\to 0}{\longrightarrow}\,\, \left.\gamma_{q_m}\star f(0,y)\right|_{y=r - \kappa}\,,
\end{equation}
imposing first a hierarchical $K$-RSB form on the matrix $\hat{q}$, taking then the $K\to\infty$ limit corresponding to fullRSB and lastly sending the number of replicas to zero. For consistency, this function must comply with a PDE, the Parisi equation, which is implemented in the above free energy via the functional Lagrange multiplier $P(t,y)$; indeed, variating with respect to $P$, we obtain
\begin{equation}
\label{eq:fullRSB_Parisi}
    \left\{\begin{aligned}
        & f(1,y) = \log \gamma_{1-q_M} \star \theta(y) \,,\\
        & \dot{f}(t,y) = -\frac{1}{2} \dot{q}(t) \left[f''(t,y) + t f'(t,y)^2\right] \qquad t_m \le t\le t_M \,.
    \end{aligned}\right.
\end{equation}
Integrating by parts the last line of Eq.~\eqref{eq:S_func} and variating with respect to $f$, we obtain a PDE for the function $P$, namely
\begin{equation}
\label{eq:fullRSB_Sompo}
    \left\{\begin{aligned}
    &P(0,y) = \gamma_{q_m}(y +\kappa - r) \equiv \rme^{-\frac{(y +\kappa - r)^2}{2q_m}}/\sqrt{2\pi q_m}\,,\\
    &\dot{P}(t,y) = \frac{\dot{q}(t)}{2}\left\{ P''(t,y) - 2 t [P(t,y)f'(t,y)]'  \right\}\,, \qquad t_m \le t \le t_M\,.
    \end{aligned}\right.
\end{equation}
Note that in these equations only the function $q(t)$ appears explicitly, but the dependence on $h(t)$ is mediated by the parameter $r$, that must be determined self-consistently on the saddle point of the free energy functional. Indeed, variations with respect to $r$ give the equation
\begin{equation}
    r = \left[\eta - \int_0^1 \diff t \,h(t)\right] \partial_r \left.\gamma_{q_m}\!\star \! f(0,y)\right|_{y=r-\kappa}\,.
    \label{eq:saddle_r}
\end{equation}
Variating with respect to $q(t)$ we obtain
\begin{multline}
    \frac{q(0) + h(0)}{4\Lambda_+(0)^2} + \frac{q(0) - h(0)}{4\Lambda_-(0)^2}  +  \int_0^t \diff u \,\frac{\dot{q}(u)+\dot{h}(u)}{4\Lambda_+(u)^2}  + \int_0^t \diff u\, \frac{\dot{q}(u)-\dot{h}(u)}{4\Lambda_-(u)^2} \\
    = \frac{\alpha}{2}\int \diff y\,P(t,y) f'(t,y)^2\,,
\label{eq:saddle_q}
\end{multline}
while variating with respect to $h(t)$ (using the saddle-point value for $r$) we can write:
\begin{multline}
    \frac{q(0) + h(0)}{4\Lambda_+(0)^2} - \frac{q(0) - h(0)}{4\Lambda_-(0)^2}  +  \int_0^t \diff u \,\frac{\dot{q}(u)+\dot{h}(u)}{4\Lambda_+(u)^2}  - \int_0^t \diff u\, \frac{\dot{q}(u)-\dot{h}(u)}{4\Lambda_-(u)^2} \\
    = \frac{\alpha}{2}  \left[\partial_r \left.\gamma_{q_m}\!\star \! f(0,y)\right|_{y=r-\kappa}\right]^2\,.
\label{eq:saddle_h}
\end{multline}
Eq.~\eqref{eq:saddle_r}--\eqref{eq:saddle_h} are the saddle point equations for the parameter $r$ and the overlap functions $q(t)$, $h(t)$.
While a general analytical solution of these equations is hopeless, useful information can be obtained with subsequent derivatives with respect to the continuous replica variable $t$. From Eq.~\eqref{eq:saddle_q} and~\eqref{eq:saddle_h}, deriving with respect to $t$ we obtain
\begin{align}
    \frac{\dot{q}(t)+\dot{h}(t)}{4\Lambda_+(t)^2}  + \frac{\dot{q}(t)-\dot{h}(t)}{4\Lambda_-(t)^2} 
    &= \frac{\alpha}{2}\dot{q}(t) \int \diff y\,P(t,y) f''(t,y)^2\,,\\
    \frac{\dot{q}(t)+\dot{h}(t)}{4\Lambda_+(t)^2}  -  \frac{\dot{q}(t)-\dot{h}(t)}{4\Lambda_-(t)^2} &= 0\,.
\end{align}
From the second one, we can solve for the ratio $\dot{q}(t)/\dot{h}(t)$,
\begin{equation}
    \frac{\dot{h}(t)}{\dot{q}(t)} = \frac{\Lambda_+(t)^2 - \Lambda_-(t)^2}{\Lambda_+(t)^2 + \Lambda_-(t)^2}\,.
\end{equation}
Substituting this result into the first one, we find
\begin{equation}
\label{eq:dot_q}
    \dot{q}(t) \left[\frac{1}{\Lambda_+(t)^2+\Lambda_-(t)^2} 
    - \frac{\alpha }{2} \int \diff y\,P(t,y) f''(t,y)^2 \right] = 0\,.
\end{equation}
This equation is equivalent to the request of \emph{marginal stability} of the fullRSB ansatz: whenever $\dot{q}(t) \neq 0$, the quantity in the square brackets must be zero on the solution of the saddle point equations. This quantity,
\begin{equation}
\label{eq:fullRSB_replicon}
    \lambda_\text{R}(t) = \frac{1}{\Lambda_+(t)^2+\Lambda_-(t)^2} 
    - \frac{\alpha }{2} \int \diff y\,P(t,y) f''(t,y)^2 \,,
\end{equation}
is called in literature \emph{replicon eigenvalue} and corresponds to the dangerous eigenvalue of the Hessian matrix $\partial^2 S/\partial q_{cd} \partial q_{ab}$ on the saddle point: it must be positive in order for the ansatz on the form of $\hat{q}$ to be stable against fluctuations in the replica space, see~\cite{dedominicisBook}.

Note that Eq.~\eqref{eq:dot_q} is already first order in $\dot{q}$. In the supposed case that replica symmetry is broken in a continuous way, it is always possible to tune the control parameters $(\eta,\kappa,\alpha)$ to be in a fullRSB region so close to the RS solution that $\dot{q}\ll 1$ in $t=t_m\approx t_M$. From Eq.~\eqref{eq:dot_q} we know that in this region $\lambda_\text{R}$ must be null at zeroth order in $\dot{q}$, that is on the RS solution: when this condition is met, we can conclude that the RS ansatz is unstable and RSB occurs~\cite{franz2017}. In the following, we will use this argument to find the critical line of instability of the RS, called de Almeida-Thouless line~\cite{dAT1978}.

\subsection{The replica symmetric solution and its stability}
\label{subsec:RS}

The replica symmetric form of the free energy can be easily obtained from the last section taking constant functions $q(t) = q_0$, $h(t) = h_0$. In this case, the free energy is
\begin{multline}
    S_{\text{RS}}[q_0,h_0,r] = \frac{1}{4} \log\!\left[\frac{(1-q_0)^2 - (\eta - h_0)^2}{1-\eta^2}\right] + \frac{q_0+h_0}{4 [1-q_0 + (\eta - h_0)]}\\
    +   \frac{q_0-h_0}{4 [1-q_0 - (\eta - h_0)]}
    - \frac{\alpha r^2}{2(\eta - h_0)} + \alpha \int \frac{\diff y \,\rme^{-\frac{y^2}{2q_0}}}{\sqrt{2\pi q_0}} \log H\!\left(\frac{\kappa - r-y}{\sqrt{1-q_0}} \right)
\,,
\label{eq:S_RS}
\end{multline}
where the function $H(x)$ is defined as in Eq.~\eqref{eq:cavity_H}. The Parisi function $f$ and its conjugate $P$ are constant in the replica variable $t$ and given by the initial and final conditions of Eq.~\eqref{eq:fullRSB_Parisi},~\eqref{eq:fullRSB_Sompo}:
\begin{equation}
    f(y) = \log \left[ H\!\left(-\frac{y}{\sqrt{1-q_0}}\right)\right]\,,\qquad P(y) = \frac{\rme^{-\frac{(y + \kappa - r)^2}{2 q_0}}}{\sqrt{2\pi q_0}}\,.
\end{equation}
At fixed value of the control parameters $\eta$, $\kappa$, $\alpha$, the saddle-point values of $q_0$, $h_0$, $r$ are obtained from Eq.~\eqref{eq:saddle_r}--\eqref{eq:saddle_h}: the saddle-point equations in this case are
\begin{subequations}
\label{eq:RS_saddle}
\begin{align}
    \frac{q_0- 2h_0 x + q_0 x^2}{(1-x^2)^2} &= \alpha F_2(\kappa - r,q_0)\,,\label{eq:RS_saddle1}\\
    \frac{h_0- 2q_0 x + h_0 x^2}{(1-x^2)^2} &= \alpha \left[F_1(\kappa - r,q_0)\right]^2\,, \label{eq:RS_saddle2}\\
    r &= x F_1(\kappa - r,q_0) \,,\label{eq:RS_saddle3}
\end{align}
\end{subequations}
where we introduced the variable
\begin{equation}
    x = \frac{\eta - h_0}{1 - q_0}
    \label{eq:MPN-corr}
\end{equation}
and defined for convenience the functions
\begin{equation}
    F_p (\kappa - r,q_0)= \int \frac{\diff y}{\sqrt{2\pi q_0}}\,\rme^{-\frac{y^2}{2q_0}} \left[\frac{\sqrt{1-q_0}\,\rme^{-\frac{(y - \kappa + r)^2}{2(1-q_0)}}}{\sqrt{2\pi}H\!\left(\frac{\kappa - r - y}{\sqrt{1-q_0}}\right)} \right]^p\,.
    \label{eq:Fp}
\end{equation}
Note that we use the same symbol to denote the quantity $x$ defined in Eq.~\eqref{eq:MPN-corr} and the one in Eq.~\eqref{eq:corr-def}, as we did with the variables $r$ in~\eqref{eq:RS_saddle} and~\eqref{eq:r-equation-C-large0}: we will see indeed in Sec.~\ref{sec:discussion} that there is a perfect correspondence between the infinite-$C$, finite-$\alpha$ MPN and the finite-$C$ OPN we discussed above. This mapping will also give us a physical interpretation for $r$ and $x$, which are variables that in the replica framework simply come out from the calculations, in the light of the cavity method we devised in Sec.~\ref{sec:cavity}.

\begin{figure}
    \centering
    \includegraphics[width=.49\textwidth]{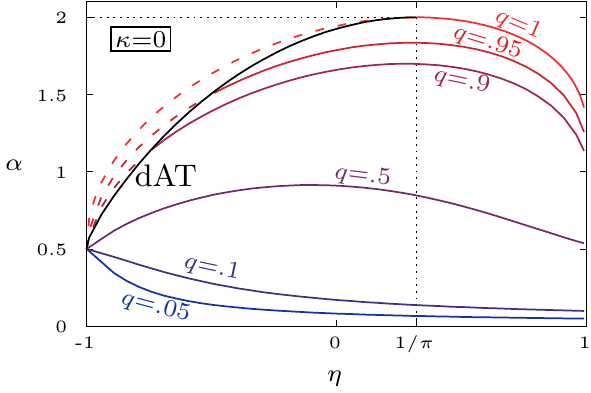}
    \includegraphics[width=.49\textwidth]{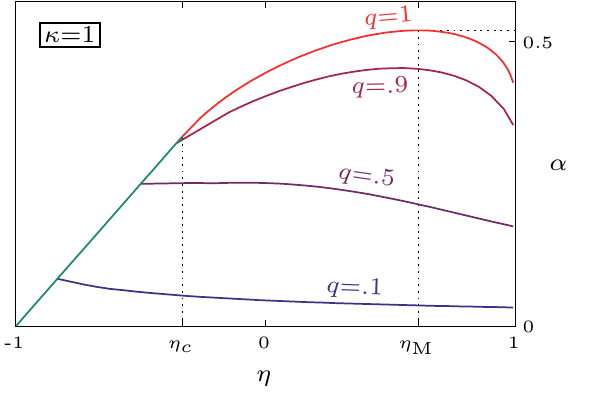}
    \caption{Phase diagram of the MPN for different values of the margin. Solutions of the RS saddle point equations at constant value of $q_0=q$ are reported. Left panel ($\kappa = 0$): the $q=1$ curve correspond to the SAT/UNSAT transition in the RS ansatz; however, below $\eta_\text{M}=1/\pi$, this curve lies above the instability de Almeida-Thouless (dAT) line, where the replicon~\eqref{eq:RS_replicon} becomes 0, signaling a RSB transition. Right panel ($\kappa = 1$): for $\eta<\eta_c(q_0=1)$ given by~\eqref{eq:replica_eta_c}, the SAT/UNSAT transition occurs on the straight line~\eqref{eq:RS_straight_line}, where $q<1$; the whole phase diagram is RS-stable.}
    \label{fig:RS_Phase diag}
\end{figure}

The system~\eqref{eq:RS_saddle} has been studied by GGY in~\cite{gardner1989phase}. As long as there is only one solution for $q_0$ at fixed values of the control parameters, it is convenient to trade $q_0$ for $\alpha$ and solve for $(\alpha,x,r)$ given $(\kappa,\eta,q_0)$. Solutions for $\kappa = 0$ and $\kappa = 1$ are plotted in Fig.~\ref{fig:RS_Phase diag}. The $\alpha(\kappa,\eta,q_0 \to 1)$ curve corresponds to the usual SAT/UNSAT transition that one finds in other CSPs, as in~\cite{monasson1999,mezard2002,krzakala2007}: increasing the number of constraints $\alpha C$, the space of solutions of the problem becomes smaller and smaller, so that different replicas of the system are more and more correlated up to the point where only a single solution is left and the overlap between replicas necessarily goes to 1. Interestingly, the maximum point $\alpha_\text{M}(\kappa)$ of this curve is not centered in $\eta=0$, but in the point $\eta_{\text{M}}(\kappa)$ where $x(\kappa,\eta_{\text{M}},q_0 \to 1) = 0$, corresponding to the critical capacity of uncorrelated spherical perceptrons, see~\cite{gardner1987perceptron}.

More crucially, for any $\kappa>0$ there is a straight line in the plane $\eta$--$\alpha$, given parametrically by
\begin{equation}
    \label{eq:RS_straight_line}
    \eta_{c} = \frac{(2 q_0 - 1) \kappa^2 - 2 q_0}{\kappa^2 + 2 q_0}\,,\qquad \alpha = \frac{q_0}{\kappa^2 + 2 q_0}\,,
\end{equation}
such that, if approached from below, $x\to -1$ and $r \to -\infty$. This line is joining the $q_0 = 1$ curve in the point
\begin{equation}
\label{eq:replica_eta_c}
    \eta_{c}(\kappa,q_0=1) = \frac{\kappa^2 - 2}{\kappa^2 + 2}\,.
\end{equation}
Solving for $q_0$ the second equation in~\eqref{eq:RS_straight_line} and substituting in the first, we obtain again Eq.~\eqref{eq:critical-line} derived for the OPN. The divergence in the variable $r$ implies that the fractional volume of solutions is going to 0, so that also this line can be interpreted as a satisfiability transition; however, on this line $q_0$ is strictly less than 1.

To explain this phenomenon, which is not present in other better-understood non-convex CSPs, at least two alternative mechanisms can be invoked:
\begin{enumerate}[(i)]
    \item The space of solutions remains connected up to the SAT/UNSAT transition, but it is not shrinking to a point: a lower-dimensional manifold of solutions with non-trivial overlaps disappears discontinuously at the transition.
    \item Before the SAT/UNSAT line an RSB transition occurs, where the space of solutions is shattered into disconnected clusters; when the SAT/UNSAT transition is approached, the size of these clusters shrinks to 0, so that only the overlaps between solutions in the same cluster go to 1; the RS approximation, which in this picture would be incorrect, interpolates between overlaps of solutions in the same and in different clusters, returning a value less than 1.
\end{enumerate}

\begin{figure}
    \centering
    \includegraphics[width=0.99\textwidth]{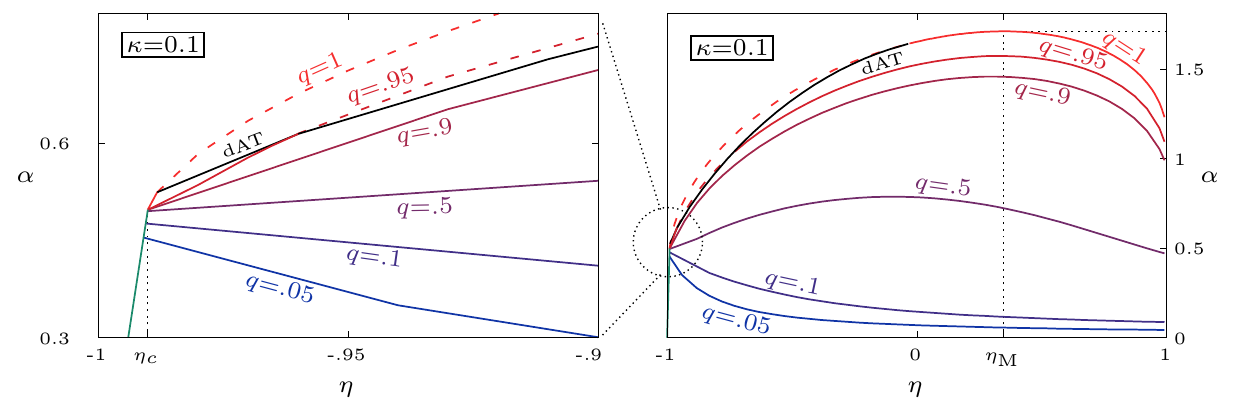}
    \caption{Phase diagram of the MPN for $\kappa=0.1$. Right: whole $\eta$--$\alpha$ plane; the dAT line is present, but the region of instability is much smaller than for $\kappa = 0$ (see Fig.~\ref{fig:RS_Phase diag}). Left: magnification of the interval $\eta\in [-1,-0.9]$: the dAT line branches from the $q_0 = 1$ line above the straight line where the SAT/UNSAT transition occurs for $q_0<1$.}
    \label{fig:RS_dAT}
\end{figure}

Thanks to the replica analysis we performed in the last section, we are able to test these hypotheses, and in particular to rule out the second scenario. Indeed, the replicon eigenvalue~\eqref{eq:fullRSB_replicon} becomes, in the RS ansatz,
\begin{equation}
\label{eq:RS_replicon}
(1-q_0)^2 \lambda_\text{R} = \frac{1}{1 +x^2} - \alpha (1-q_0)^2 \int \diff y \, \frac{\rme^{-\frac{(y +\kappa - r)^2}{2 q_0}}}{\sqrt{2\pi q_0}} \left\{\frac{\diff^2}{\diff y^2} \log \left[H\!\left(-\frac{y}{\sqrt{1-q_0}} \right) \right]\right\}^2\,,
\end{equation}
to be evaluated on the solution of the system~\eqref{eq:RS_saddle}: where this quantity turns negative, RSB occurs. We plot the corresponding $\lambda_{\text{R}} = 0$ curve (de Almeida-Thouless line, dAT) in Fig.~\ref{fig:RS_Phase diag} for $\kappa = 0$ (left): it starts from $\eta = -1$, $\alpha = 1/2$ and reaches the point $(\eta_\text{M}(0),\alpha_\text{M}(0))$, cutting a slab of the phase diagram where there is indeed a clustering transition in the space of solutions.\footnote{The determination of the exact nature of this transition in terms of the RSB structure is beyond the scope of this paper.} However, this region of instability becomes smaller and smaller increasing $\kappa$, up to disappearing completely at around $\kappa \approx 0.28$. For $\kappa = 1$, the whole phase diagram in Fig.~\ref{fig:RS_Phase diag} (right) is in the RS-stable region. We report the intermediate case of $\kappa = 0.1$ in Fig.~\ref{fig:RS_dAT}: from the inset around the point $\eta_c(0.1)$ (left) we can see that the dAT curve starts strictly from the right of the straight line where the SAT/UNSAT transition occurs for $q_0<1$. This fact can also be checked analytically: from \eqref{eq:RS_straight_line}, $\alpha<1/2$ on the line, so the replicon is strictly positive (the coefficient of $\alpha$ in Eq.~\eqref{eq:RS_replicon} is exactly $-1$ when $r\to-\infty$, while the first term is $1/2$ for $x\to -1$). We can thus conclude that an instability of the RS solution is present only for a limited range of the control parameters and is not responsible for the satisfiability transition at $q_0<1$.

In the next section we will discuss the strict analogy between OPN and MPN we analyzed so far and we will argue on the  common nature of their phase transitions.

\section{Discussion \label{sec:discussion}}

In this paper we have explored the questions asked by GGY in the conclusions of \cite{gardner1989phase}. Our analysis suggests that the existence of critical $\eta_c$ is in fact \emph{not} related to a clustering transition described by RSB, since the replicon eigenvalue is always positive in this region, and that it is related instead to a full anti-correlation between variables. We arrive at this picture by mapping our problem into a minimal model with very similar features, the soft OPN. This model suggests a physical interpretation for the order parameters $x$ and $r$ found by GGY for the MPN. We have intentionally used the same notation for both in sections \ref{sec:cavity} and \ref{sec:replica} to highlight the analogy.

In the OPN, $x$ corresponds to the correlation coefficient between $J_{ij}$ and $J_{ji}$. In the MPN, it was introduced as the specific combination of $\eta$, $q_0$, and $h_0$ in (\ref{eq:MPN-corr}). If one interprets the integrand in (\ref{eq:volume}) as a probability distribution analogous to the one of (\ref{eq:opn-model}) but with a hard asymmetry constraint, then we can also introduce the notion of average $\expected{\cdot}$ for the MPN, similar to the one introduced in \cite{mezard1989space}. Under this point of view then $x$ in both models only differs by the average over the disorder present in the MPN.
\begin{equation}
    \begin{aligned}
    \textrm{OPN:} \quad & x  = \frac{\expected{J_{ij}J_{ji}} - \expected{J_{ij}} \expected{J_{ji}}}{\expected{J_{ij}^2} - \expected{J_{ij}}^2} \,;\\
    \textrm{MPN:} \quad & x  = \frac{\eta - h}{1 - q} = \frac{\mean{\expected{J_{ij}J_{ji}}} - \mean{\expected{J_{ij}} \expected{J_{ji}}}}{ \mean{\expected{J_{ij}^2}} - \mean{\expected{J_{ij}}^2}} \,.
\end{aligned}
\end{equation}
The first corroboration of this interpretation comes from the fact that at $x=0$ one recovers the theory developed in \cite{gardner1987perceptron} for uncorrelated perceptrons. Additionally, it also varies from $x = -1$ at the critical value of asymmetry, $\eta_c$, to $x=1$ for $\eta = 1$, in the same way as in the OPN.

The other related observable is $r$. In the OPN it corresponds to $\sqrt{C}\mu$, which can also be defined as the expected value of the gap (before truncation at $\kappa$), as seen in (\ref{eq:gap-distribution}), 
\begin{align}
    \text{OPN:}\quad r ={}& \expected{\frac{1}{\sqrt{C}}\sum_{j\in \partial i}J_{ij}}_{\nu_{\partial i \to i}} = \sqrt{C} \expected{J_{ij}}_{\nu_{ij \to i}} \,,
\end{align}
where the expected value $\expected{\cdot}_{\nu_{\partial i \to i}}$ means all $J_{ij}$'s in that sum are sampled from their corresponding $\nu_{{ij}\to i}$. For the analogous expression in the MPN we need to include a limit of $C\to\infty$ and the presence of the quenched patterns in the field,
\begin{align}
    \text{MPN:}\quad r =\lim_{C\to\infty} \mean{\expected{{\frac{1}{\sqrt{C}}\sum_{j\in\partial i}\xi_i^\mu J_{ij} \xi_j^\mu}}_{\nu_{\partial i \to i}}} .
\end{align}
If we also look at the large $C$ limit of the OPN, we can see that both equations have basically the same form, except for the average over the disorder. Using $\calN(J,J_0|\bSigma)$ to denote a zero-mean Gaussian with covariance $\bSigma$, we can write both equations for r in each model, (\ref{eq:r-equation-C-large}) and (\ref{eq:RS_saddle}c), in the following way,
\begin{equation}
    \begin{aligned}
    \text{OPN:}\quad\qquad  r ={} & \frac{\int\rmd J\rmd J_0\, J \, \calN(J,J_0|\bSigma^\text{O}) \theta(J_0 + r - \kappa)}{\int\rmd J\rmd J_0 \, \calN(J,J_0|\bSigma^\text{O}) \theta(J_0 + r - \kappa)} \,,\\
   \bSigma^\text{O} ={} & \p{\begin{array}{cc}
    1     & \eta \\
    \eta     & 1
    \end{array}} ;\\
    \text{MPN:}\quad\qquad r ={} & \int\frac{\rmd t}{\sqrt{2\pi q_0}}\rme^{-\frac{1}{2}\frac{(t - r)^2}{q_0}}\frac{\int\rmd J\rmd J_0\, J \, \calN(J,J_0|\bSigma^\text{M}) \theta(J_0 + t - \kappa)}{\int\rmd J\rmd J_0 \, \calN(J,J_0|\bSigma^\text{M}) \theta(J_0 + t - \kappa)} \,,\\
    \bSigma^\text{M} ={} & \p{\begin{array}{cc}
    1 - q_0     & \eta - h_0 \\
    \eta - h_0     & 1 - q_0
    \end{array}}.
\end{aligned}
\end{equation}

It is clear then that the role of the average over disorder corresponds to the first Gaussian integration over $t$. Such interpretation is actually shown for the case of the uncorrelated perceptron in \cite{mezard1989space}. It is important to point out that the nature of $r$ is directly related to the asymmetry constraints. As a matter of fact, for the uncorrelated perceptrons with $x = 0$ we have $r=0$. As it can be seen in both models, $r$ plays the role of a self-induced margin. The origin is clear with the cavity interpretation: if one wants to add a new node $i$ to a model where it was not present before, one needs not only to solve the problem of choosing the $J_{ij}$'s to satisfy the constraint over $i$, one needs also the new $J_{ji}$ correlated with $J_{ij}$ to surpass $\kappa$ and the rest of the elements in the constraint over node $j$. The correlation between variables is making the problem harder, and therefore reducing the volume of solutions. 

Thus, we can provide a picture for the SAT/UNSAT transition observed at $\eta_c$ with $q_0<1$ for the MPN that does not require RSB to be understood intuitively. We conclude that it should be the same behaviour as the one described for the OPN at criticality in Section \ref{subsec:asympt_lambda}. In this way the divergence $r\to-\infty$ in the MPN actually has a physical interpretation: it corresponds to the point where the frustration due to the asymmetry constraint explodes as a consequence of the perfect anti-correlation between $J_{ij}$ and $J_{ji}$. At this point a connected volume of solutions still exists, but it is concentrated in a measure-zero set. Therefore, in this critical line of the MPN we have an extended set of solutions with variance $1-q_0$ concentrated in lower dimensional manifold. In this way we can have simultaneously a zero volume of solutions and $q_0<1$. Even more, this transition survives all the way down to $\alpha \approx 0$, as we have shown that it is present even in the one pattern case with $C\to\infty$. 

To conclude, the present paper contributes to explain an alternative mechanism for a random CSP with continuous variables to exhibit a SAT/UNSAT transition, which was not understood before in the general framework of disordered systems~\cite{krzakala2007,franz2017}. Moreover, it confirms that the results of~\cite{gardner1989phase} on the storage capacity of asymmetric recurrent neural networks were correctly describing this new phenomenon rather than being an artifact of the replica symmetric approach.

In the future, it would be interesting to derive GGY's equations for $q_0,h_0$ and $r$, in the MPN \eqref{eq:RS_saddle} directly with the cavity method. The problem is nonstandard as the constraints at the factor nodes are correlated with each other, the reason being that $\xi^\mu_i$ appears both with $J_{ij}$ and with $J_{ji}$ in two different constraints. This deviates from the standard theory of random graphical models that assumes factors are independent random variables. Establishing the correct connection with the cavity method should shed light on the one assumption from GGY's calculation we have not explored in this paper, the mean-field like assumption \eqref{eq:mean-field-assumption}. The very strong resemblance between the OPN and the MPN could mean that this assumption is too strong. Understanding the exact physical meaning of this assumption within the cavity should answer this question. This is left for future work.

Another very promising research line for future investigation is to determine the behaviour of the fully connected MPN, for which $C\sim N$: in this case, the replica analysis performed by GGY should not be applied in a straightforward way (see~\cite{gardner1989phase,theumann1996} and point (i) in the Introduction of the present paper). If this is indeed the case, significant deviations from the standard mean field theory could be observed, as already suggested numerically by \cite{gardner1989phase,ventura2021}.

\section*{Acknowledgments}

All the authors are supported by a grant from the Simons foundation (grant No. 454941, S. Franz). SF is a member of the Institut Universitaire de France. The authors would like to thank Enrico Ventura, Francesco Zamponi, Louise Budzynski and Valentina Ros for discussions and suggestions.

\printbibliography

\appendix

\section{One pattern network with replicas \label{app:opn-replica}}

We calculate volume of solutions for the one pattern model in the limit $C \to \infty$ \emph{after} $N \to \infty$. We do the calculation directly with replicas \emph{without} doing the mean-field assumption \eqref{eq:mean-field-assumption} of GGY used in \cite{gardner1989phase}. We start from the fractional synaptic volume

\begin{align}
    V = {}& \int\rmd\bJ \rho_0(\bJ) \prod_{i}  \theta\p{\sum_j A_{ij}\frac{J_{ij}}{\sqrt{C}} - \kappa}
    ,
\end{align}
where we assume $\bA$ is a random graph with connectivity $C$ and that $\rho_0(\bJ)$ enforces the desired asymmetry constraints. 
We would like to calculate
\begin{align}
    \frac{1}{N}\expected{\log V} = \lim_{n\to 0} \frac{1}{Nn}\log \expected{V^n}\,.
\end{align}

In this case $\expected{\cdot}$ means averaging over the graphs, $\bA$, and the weights, $\bJ$. We choose Erd\"os-R\'enyi random graphs with connectivity $C$, this simplifies the calculation while keeping the same behaviour at $C\to\infty$:
\begin{align}
    p(\bA) = \prod_{i<j}\p{\frac{C}{N}\delta_{A_{ij},1} + \p{1 - \frac{C}{N}}\delta_{A_{ij},0}}\,.
\end{align}
For $\rho_0(\bJ)$ we choose a product of $\eta$ correlated Gaussian bivariate distributions.
\begin{align}
    \rho_0(\bJ) = \prod_{i<j} \calN_{\eta}(J_{ij},J_{ji})\,.
\end{align}

Using the integral representation of the $\theta(x)$, we now perform the replica calculation in a similar way to \cite{kuhn2008spectra,lopez2020imaginary}. In this way avoid doing any unwanted assumptions. We begin by rearranging in such a way that all averages can be done explicitly.
\begin{align}
    \expected{V^n} = &\expected{ \int \prod_a \rmd \bJ^a \rho_0(\bJ^a) \cdot \prod_{ia} \theta\p{\sum_j A_{ij}\frac{J_{ij}^a}{\sqrt{C}} - \kappa}}\nonumber\\
     = & \int \prod_a \rmd \bJ^a \rho_0(\bJ^a) \int_\kappa^\infty \rmd \dblambda \int_{-\infty}^\infty \frac{\rmd \dbx}{(2\pi)^{Nn}}\rme^{\rmi \dblambda\cdot\dbx }\expected{\rme^{-\rmi \sum_{ij} A_{ij} \sum_a x_i^a \frac{J_{ij}^a}{\sqrt{C}} }} \nonumber\\
     = & \int_\kappa^\infty \rmd \dblambda \int_{-\infty}^\infty \frac{\rmd \dbx}{(2\pi)^{Nn}}\rme^{\rmi \dblambda\cdot\dbx } \exp\p{ \frac{C}{2N} \sum_{ij} 
     \prod_a \expected{\rme^{-ix_i^a J/\sqrt{C} - \rmi x_j^a J'/\sqrt{C}}}_{\calN_\eta (J,J')} - 1} \nonumber\\
     = & \int_\kappa^\infty \rmd \dblambda \int_{-\infty}^\infty \frac{\rmd \dbx}{(2\pi)^{Nn}}\rme^{\rmi \dblambda\cdot\dbx + \frac{C}{2N}\sum_{ij} (\rme^{-\frac{1}{2C}\bx^i\cdot\bx^i- \frac{\eta}{C} \bx^i\cdot\bx^j - \frac{1}{2C}\bx^j\cdot\bx^j} -1)}
\end{align}
We now perform a change of variable of the integral by noticing all the terms of the integrand can be rewritten as integrals over the next distribution over $n$ dimensional vectors, $\bx$. 
\begin{align}
    P(\bx) = {}& \frac{1}{N}\sum_{i=1}^N \delta(\bx - \bx^i) 
\end{align}
We enforce this definition using an infinite product of delta functions inserted in the integral as a number one,
\begin{align}
    1 = \int\calD P \calD \hP \rme^{N \rmi \int\rmd\bx P(\bx) \hP(\bx) - \rmi \sum_{i=1}^N \hP(\bx^i)}.
\end{align}
Inserting this one in the integral we get
\begin{align}
    \expected{V^n} = & \int\calD P \calD \hP \; \rme^{ N S[P,\hP]}
\end{align}
Where we have defined
\begin{align}
    S[P,\hP] = {}& \rmi \int \rmd \bx P(\bx) \hP(\bx) + \frac{C}{2} \int \rmd \bx \rmd \bx' P(\bx) U(\bx,\bx') P(\bx') \nonumber\\
    & + \log \int_{\kappa}^{\infty} \rmd\blambda \int_{-\infty}^\infty \frac{\rmd \bx}{(2\pi)^n} \rme^{\rmi\blambda\cdot\bx - \rmi \hP(\bx)},\nonumber\\
    U(\bx,\bx') = {}& \rme^{-\frac{1}{2C}\bx\cdot\bx -\frac{\eta}{C} \bx\cdot\bx' - \frac{1}{2C}\bx'\cdot\bx'} - 1\nonumber
\end{align}
This integral can now be evaluated for large $N$ with the saddle point method. We get the next saddle point equations by looking at the stationary point of $S[P,\hP]$ we will denote by $P_s$ and $\hP_S$,
\begin{align}
    P_s(\bx) = {}& \frac{1}{\calZ} \int_{\kappa}^{\infty} \frac{\rmd \blambda}{(2\pi)^n} \rme^{\rmi\blambda\cdot\bx - \rmi \hP_s(\bx)}, \nonumber\\
    -\rmi \hP_s(\bx) = {}& C\int\rmd\bx' U(\bx,\bx') P_s(\bx').
\end{align}
We are now interested in taking the limit $C\to\infty$, this turns our saddle point equations into
\begin{align}
    P_s(\bx) = {}& \frac{1}{\calZ} \int_{\kappa}^{\infty} \frac{\rmd \blambda}{(2\pi)^n} \rme^{\rmi\blambda\cdot\bx - \rmi \hP_s(\bx)} \nonumber\\
    -\rmi \hP_s(\bx) = {}& -\frac{1}{2}\bx\cdot\bx -\eta \bx \cdot \int \rmd\bx' \bx' P_s(\bx') - \frac{1}{2}\int \rmd \bx' \bx'\cdot \bx' P_s(\bx')
\end{align}
Notice that taking the limit $C\to \infty$ after making the saddle point approximation implies we have taken this limit after $N\to \infty$. This corresponds to being on the so called diluted limit which can be achieved by taking $C  \sim \Order{\log N}$ for example. In this regime we get the next form for $P_s$,
\begin{align}
    P_s(\bx) = {}& \frac{1}{Z} \int_{\kappa}^{\infty} \frac{\rmd \blambda}{(2\pi)^n} \rme^{\rmi\blambda\cdot\bx -\frac{1}{2} \bx\cdot\bx -\rmi  \bx\cdot \br}\nonumber\\
    Z = {}&  \prod_a \int_{-\infty}^\infty \rmd t \,\rme^{-\frac{1}{2}t^2} \theta(r_a +t - \kappa)
\end{align}
Where $\br$ has to be determined from the equation, 
\begin{align}
    \rmi \br = \eta \int \rmd\bx \, \bx P_s(\bx) = \eta \rmi\nabla_{\br} \log Z. 
\end{align}
If we now assume replica symmetry, $r_a = r$, we get the equation,
\begin{align}
    r  = \eta \frac{ \rme^{-\frac{1}{2}(r - \kappa)^2}}{\int_{\kappa - r}^\infty \rmd t \,\rme^{-\frac{1}{2}t^2} }
\end{align}

\section{Details of the replica calculations for the MPN
\label{app:replica}}
In this section we report for quick reference some of the steps needed to derive the equations reported in Sec.~\ref{sec:replica}, following~\cite{gardner1989phase}. For a more in-depth discussion, we address the interested reader to the GGY paper.
Introducing via a delta function the gap variables
\begin{equation}
\label{eq:app_rep_gaps}
\Delta^{a}_i(J) = \xi_i \sum_{j\in \partial i} \frac{J_{ij}^a}{\sqrt{C}}\, \xi_j \,,
\end{equation}
we can write the $\theta$-constraints of the replicated problem as
\begin{equation}
\prod_{i,a}\theta \Biggl(\xi_i \sum_{j \in \partial i} \frac{J_{ij}^a}{\sqrt{C}}\, \xi_j - \kappa \Biggr)
 = \int \prod_{i,a}\frac{\diff x^{a}_i \diff \Delta^{a}_i \theta(\Delta^{a}_i - \kappa)}{2\pi}\, \rme^{i \sum_{i,a}x^{a}_i \Delta^{a}_i   - i\sum_{i,a}x^{a}_i \xi_i \sum_{j\in \partial i} \frac{J_{ij}^a}{\sqrt{C}}\, \xi_j }\,.
\end{equation}
A crucial hypothesis in the GGY paper is that the variable
\begin{equation}
    \sum_{i,a}x^{a}_i \xi_i \sum_{j\in \partial i} \frac{J_{ij}^a}{\sqrt{C}}\, \xi_j
\end{equation}
is Gaussian with respect to the probability distribution of the patterns. This hypothesis is justified at least in the diluted limit $C\sim \log N$, even though at least one reference~\cite{theumann1996} points out that this restriction could be over-pessimistic. Given that, its mean is zero and its variance is
\begin{equation}
\sum_{i,k,a,b}x^a_i x^b_k \sum_{\substack{j\in\partial i\\l \in \partial k}} \frac{J_{ij}^a J_{kl}^b}{C}\, \mean{\xi_i \xi_k \xi_j  \xi_l} =\sum_{a,b} \sum_{i} x^a_i x^b_i \sum_{j\in \partial i}\frac{J_{ij}^a J_{ij}^b}{C}\, + \sum_{a,b} \sum_{i} x^a_i  \sum_{j\in\partial i} x^b_j\frac{J_{ij}^a J_{ji}^b}{C}\,.
\end{equation}
To evaluate the second term, a \emph{mean-field}, site-symmetric assumption is taken: requiring that the internal sum over $j\in \partial i$ does not depend on $i$, we can substitute it with its average over the sites:
\begin{equation}
\sum_{j\in \partial i}  x^b_j\frac{J_{ij}^a J_{ji}^b}{C} \to \frac{1}{N}\sum_{k,j\in \partial k}  x^b_j\frac{J_{kj}^a J_{jk}^b}{C} = \frac{1}{N}\sum_{j}  x^b_j\sum_{k\in \partial j}\frac{J_{kj}^a J_{jk}^b}{C} = \frac{1}{N}\sum_{j}  x^b_jh^j_{ab}
\end{equation}
Now both terms can be written in terms of global replica overlaps ${\hat{q}}^i$ and ${\hat{h}}^i$. With the additional site-symmetric assumption that these overlaps do not depend on $i$, we arrive at the form
\begin{multline}
\mean{\prod_{\mu,i,a}\theta \Biggl(\xi_i^\mu \sum_{j\neq i} \frac{J_{ij}^a}{\sqrt{C}}\, \xi_j^\mu - \kappa \Biggr)} \\
= \left\{\int \prod_{i,a}\frac{\diff x^a_i \diff \Delta^a_i \theta(\Delta^a_i - \kappa)}{2\pi} \,\rme^{i \sum_{i,a}x^a_i \Delta^a_i - \frac{1}{2}\sum_{i,a,b}q_{ab} x^a_i x^b_i - \frac{1}{2N}\sum_{i,j,a,b}h_{ab} x^a_i x^b_j}\right\}^{\alpha C}.
\end{multline}
To linearize the double summation over the sites indices in the last term at the exponent and factorize the exponential over the sites $i$, we can perform an Hubbard-Stratonovich transformation, obtaining
\begin{multline}
\Biggl\{\int\frac{ \prod_{a} \diff r^a}{\sqrt{(2\pi)^n \det h}}\Biggl[\int \prod_{a}\frac{\diff x^a \diff \Delta^a \theta(\Delta^a)}{2\pi} \\
\times \rme^{i \sum_{a} x^a \Delta^a - i \sum_{a}(r^a - \kappa)x^a  - \frac{1}{2}\sum_{a,b}q_{ab} x^a x^b - \frac{1}{2}\sum_{a,b}h^{-1}_{ab} r^a r^b}\Biggr]^N\Biggr\}^{\alpha C}.
\end{multline}
Using the identity
\begin{equation}
\rme^{-i \sum_{a}(r^a - \kappa)x^a  - \frac{1}{2}\sum_{a,b}q_{ab} x^a x^b} = \left.\rme^{\frac{1}{2}\sum_{a,b}q_{ab} \frac{\partial^2}{\partial \nu^a \partial \nu^b}} \rme^{-i\sum_a \nu^a x^a}\right|_{\nu^a =  r^a -\kappa}\,,
\end{equation}
we obtain Eq.~\eqref{eq:replica_freeEnergy0} in the main text.

{

\section{Numerical methods}
\label{sec:numerics}
\subsection{MCMC sampling}

We show the algorithm used to obtained samples from \eqref{eq:opn-model}. For a working implementation of the algorithm go to \url{https://github.com/aguirreFabian/asymmetric-networks}.

In order to generate samples from the OPN ensemble \eqref{eq:opn-model}, one needs to be able to control two different kinds of biases, the one given by the asymmetry and the one given by the hard row constraints. The method of choice was Markov Chain Monte Carlo sampling, \cite{landau2021guide}. In order to find the appropriate sector of the space of $\bJ$'s that contains typical samples of \eqref{eq:opn-model}, we need to \emph{anneal} the system into it. Trying to directly sample from \eqref{eq:opn-model} is in theory possible with MCMC, but the algorithm is likely to get stuck and have very long relaxation times. To get around this problem, we begin by biasing at first with respect to a different distribution, where the row constraints are soft and have a tuneable strength. We define this distribution as $\rho_S(\bJ|\gamma)$
\begin{subequations}
\label{eq:opn-soft-annealing}
\begin{align}
    \rho_S(\bJ|\gamma)  ={} & \frac{1}{Z_{S}}\prod_{(i,j) \in E} \calN_{\lambda}(\bJ_{ij}) \prod_{i \in V} \rme^{ - \gamma C (\Delta_i - \kappa )^2 \theta(\kappa - \Delta_i)} = \frac{1}{Z_S} \rme^{\calH_\lambda(\bJ) + \gamma \calH_\calC (\bJ)} \label{eq:opn-soft-annealing-model}\\
    \calH_\lambda(\bJ) ={} & \sum_{(i,j) \in E} \log \calN_{\lambda}(\bJ_{ij})  = - \sum_{(i,j)\in E} \frac{1}{2 (1-\lambda^2)} (J_{ij}^2 + J_{ji}^2 - 2 \lambda J_{ij}J_{ji})\label{eq:opn-soft-annealing-asymmetryl}\\
    \calH_\calC (\bJ) ={} & - \sum_{i \in V} (\sum_{j\in \partial i} J_{ij} - \kappa\sqrt{C})^2 \theta(\kappa\sqrt{C} - \sum_{j\in \partial i} J_{ij}) \label{eq:opn-soft-annealing-constraints}
\end{align}
\end{subequations}
In this distribution $\bJ$'s that do not satisfy the pattern constraints are allowed, but they are penalized. The parameter $\gamma$ tunes the strength of the penalty in such a way that at $\gamma \to \infty$ we recover the original model \eqref{eq:opn-model}, $\rho_S(\bJ|\gamma) \underset{\gamma \to \infty}{\longrightarrow} \rho(\bJ)$. Using this property, we start our algorithm by using MCMC to sample from \eqref{eq:opn-soft-annealing-model} for increasing values of $\gamma$. In practice, we set a sequence of length $P$ of values of $\gamma$, $\{\gamma_p\}_{p = 1,\dots,P}$, and for each value of $\gamma_p$ we run $T_0$ MCMC steps. When $\gamma_p$ is small most of the row constraints are unsatisfied (UNSAT), but as $\gamma_p$ increases some of them start becoming satisfied (SAT). Once the maximum value of $\gamma_P = \gamma_{\max}$ is reached, it is useful to do some extra MCMC steps until the number of UNSAT constraints stabilizes in a certain value. Even though most of the constraints may remain UNSAT, this initial annealing is good enough to then get good samples of \eqref{eq:opn-model}. To make all constraints SAT, we simply force each constraint by pushing a random synapse $J_{ij}$ to $\kappa \sqrt{C} - \sum_{k \in \partial i \setminus j} J_{ik}$ while completely disregarding the asymmetry bias given by $\lambda$. After this procedure we get a $\bJ_0$ that satisfies all the row constraints and that hopefully has a remnant of the previous asymmetry bias. We then use this $\bJ_0$ as a starting point for the true MCMC of \eqref{eq:opn-model} and perform $T$ MCMC steps until the chain converges. See Algorithm \ref{alg:MCMC}.

\begin{algorithm}
\caption{ }\label{alg:MCMC}
\begin{algorithmic}
\Require Graph, $G = (V,E)$; asymmetry, $\lambda$; margin, $\kappa$; annealing rounds, $P$; annealing steps, $T_0$; MCMC steps, T; typical move sizes, $s_1$ and $s_2$; interval of annealing strengths, $[\gamma_{\min} ,\gamma_{\max}]$

\Ensure values of synapses $\bJ$ sample of \eqref{eq:opn-model}

\State $\bJ \gets$ random graph with asymmetry $\lambda$
\State $\{\gamma_p\} \gets$ sequence of annealing strengths, $\gamma_{\min} = \gamma_1\le\dots\le\gamma_p\le\dots\le\gamma_P = \gamma_{\max}$ 

\For{$p=1$ to $P$}\Comment{Annealing}
\For{$t = 1$ to $T_0$}
\State Choose $(i,j) \in E$ at random and one component $J_e \in \{J_{ij}, J_{ji}\}$ at random 
\State Propose a new $\bJ^*$ which differs only by $J_e^* \gets J_e + \delta s_1$, with $\delta\sim\calN(0,1)$
\State With probability $p = \min\{1, \rme^{\Delta \calH_{\lambda} + \gamma_p \Delta \calH_\calC}\}$ update $\bJ \gets \bJ^*$
\EndFor
\EndFor 

\For{$i\in V$}\Comment{Force patterns}
\If{$\sum_j J_{ij}<\kappa \sqrt{C}$}
\State Choose $j \in \partial i$ at random
\State $J_{ij} \gets \kappa \sqrt{C} - \sum_{k \in \partial i\setminus j} J_{ik}$
\EndIf
\EndFor

\For{$t = 1$ to $T$}\Comment{MCMC sampling}

\State Choose $(i,j) \in E$ at random and one component $J_e \in \{J_{ij}, J_{ji}\}$ at random 
\State Propose a new $\bJ^*$ which differs only by $J_e^* \gets J_e + \delta s_2$, with $\delta \sim \calN(0,1)$
\If {$\rho(\bJ^*)>0$ (\emph{i.e.} all constraints are satisfied)}
\State With probability $p = \min\{1, \rme^{\Delta \calH_{\lambda}} \}$ update $\bJ \gets \bJ^*$
\EndIf
\EndFor

\Return{$\bJ$}
\end{algorithmic}
\end{algorithm}

\subsection{Numerical solution of the RS saddle-point system for the MPN}
The way to solve Eq.~\eqref{eq:RS_saddle} is detailed in the original reference~\cite{gardner1989phase} by GGY and here in the main text. We can easily solve for $(\alpha,x,r)$ given $(\kappa,\eta,q_0)$ by iteration. The only slightly non-trivial observation we did to speed up the algorithm is that the function in the square bracket of Eq.~\eqref{eq:Fp} approaches very fast its asymptotic form for large value of its argument, so we used the representation
\begin{equation}
\frac{\sqrt{1-q_0}e^{-\frac{y^2}{2(1-q_0)}}}{\sqrt{2\pi}H(-\frac{y}{\sqrt{1-q_0}})} \sim \begin{cases}
-y - \frac{1-q_0}{y}		
& \text{if $ y < - s \sqrt{1-q_0}$}\\
\frac{\sqrt{1-q_0}e^{-\frac{y^2}{2(1-q_0)}}}{\sqrt{2\pi}H(-\frac{y}{\sqrt{1-q_0}})}
& \text{if $- s \sqrt{1-q_0} < y <  s \sqrt{1-q_0}$}\\
0
& \text{if $ y >  s \sqrt{1-q_0}$}
\end{cases}
\end{equation}
where $s$ is a certain scaling cut-off (we used $s=5$ for the figures in the main text, observing almost no difference increasing this value). We include a Wolfram Mathematica notebook to obtain our figures in the repository.

}

\end{document}